\documentclass[longauth]{aa}

\usepackage{graphicx}
\usepackage{natbib}
\usepackage{scalerel}
\usepackage{txfonts}
\usepackage[table]{xcolor}
\usepackage{hyperref}
\hypersetup{
    colorlinks=true,
    linkcolor=blue,
    filecolor=magenta,      
    urlcolor=blue,
    citecolor=blue
}
\makeatletter
\renewcommand*\aa@pageof{, page \thepage{} of \pageref*{LastPage}}
\makeatother

\usepackage[utf8]{inputenc}
\usepackage{euclid}
\usepackage[switch, modulo]{lineno}

\begin{document}

\newcommand{\zsvalue}{0.4058} 
\newcommand{\lensname}{Altieri's lens}
\newcommand{\galaxyname}{NGC~6505}
\newcommand{\galaxycoords}{RA~\ra{17;51;07.46}, Dec~$+\ang{65;31}\ang{;;50.78}$}

\newcommand{\zl}{z_\mathrm{d}}
\newcommand{\zs}{z_\mathrm{s}}
\newcommand{\erad}{\theta_\mathrm{E}}
\newcommand{\sigmav}{\sigma_\mathrm{v}}
\newcommand{\dmfraction}{f_\mathrm{DM}}
\newcommand{\Dls}{D_\mathrm{ls}}
\newcommand{\Dl}{D_\mathrm{l}}
\newcommand{\etheta}{\theta_q}
\newcommand{\lensx}{\theta_{\mathrm{l}_x}}
\newcommand{\lensy}{\theta_{\mathrm{l}_y}}
\newcommand{\posa}{\phi_\mathrm{L}}
\newcommand{\lambdas}{\lambda_\mathrm{s}}
\newcommand{\extshear}{\gamma_\mathrm{ext}}
\newcommand{\imfmismatch}{\alpha_\mathrm{IMF}}
\newcommand{\Reff}{R_\mathrm{eff}}

\newcommand{\chisqnu}{\chi^2_\nu}
\newcommand{\ndof}{N_\mathrm{dof}}
\newcommand\given[1][]{\:#1\vert\:}	
\newcommand{\prob}[2]{\mathrm{Pr}\left(#1\vert#2\right)}
\newcommand{\prior}[1]{\mathrm{Pr}\left(#1\right)}
\newcommand{\ev}{\varepsilon}

\newcommand{\Msun}{M_\odot}

\title{
    \Euclid\/: A complete Einstein ring in NGC 6505
    \thanks{This paper is published on behalf of the Euclid Consortium}
}

   

\newcommand{\orcid}[1]{} 
\author{C.~M.~O'Riordan\orcid{0000-0003-2227-1998}\thanks{\email{conor@mpa-garching.mpg.de}}\inst{\ref{aff1}}
\and L.~J.~Oldham\inst{\ref{aff2}}
\and A.~Nersesian\orcid{0000-0001-6843-409X}\inst{\ref{aff3},\ref{aff4}}
\and T.~Li\orcid{0009-0005-5008-0381}\inst{\ref{aff2}}
\and T.~E.~Collett\orcid{0000-0001-5564-3140}\inst{\ref{aff2}}
\and D.~Sluse\orcid{0000-0001-6116-2095}\inst{\ref{aff3}}
\and B.~Altieri\orcid{0000-0003-3936-0284}\inst{\ref{aff5}}
\and B.~Cl\'ement\orcid{0000-0002-7966-3661}\inst{\ref{aff6},\ref{aff7}}
\and K.~Vasan~G.~C.~\orcid{0000-0002-2645-679X}\inst{\ref{aff8}}
\and S.~Rhoades\orcid{0009-0007-0184-8176}\inst{\ref{aff8}}
\and Y.~Chen\orcid{0000-0003-4520-5395}\inst{\ref{aff8}}
\and T.~Jones\orcid{0000-0001-5860-3419}\inst{\ref{aff8}}
\and C.~Adami\orcid{0000-0002-8904-3925}\inst{\ref{aff9}}
\and R.~Gavazzi\orcid{0000-0002-5540-6935}\inst{\ref{aff9},\ref{aff10}}
\and S.~Vegetti\orcid{0009-0006-0592-2882}\inst{\ref{aff1}}
\and D.~M.~Powell\orcid{0000-0002-4912-9943}\inst{\ref{aff1}}
\and J.~A.~Acevedo~Barroso\orcid{0000-0002-9654-1711}\inst{\ref{aff6}}
\and I.~T.~Andika\orcid{0000-0001-6102-9526}\inst{\ref{aff11},\ref{aff1}}
\and R.~Bhatawdekar\orcid{0000-0003-0883-2226}\inst{\ref{aff5}}
\and A.~R.~Cooray\orcid{0000-0002-3892-0190}\inst{\ref{aff12}}
\and G.~Despali\orcid{0000-0001-6150-4112}\inst{\ref{aff13},\ref{aff14},\ref{aff15}}
\and J.~M.~Diego\orcid{0000-0001-9065-3926}\inst{\ref{aff16}}
\and L.~R.~Ecker\inst{\ref{aff17},\ref{aff18}}
\and A.~Galan\orcid{0000-0003-2547-9815}\inst{\ref{aff11},\ref{aff1}}
\and P.~G\'omez-Alvarez\orcid{0000-0002-8594-5358}\inst{\ref{aff19},\ref{aff5}}
\and L.~Leuzzi\orcid{0009-0006-4479-7017}\inst{\ref{aff13},\ref{aff14}}
\and M.~Meneghetti\orcid{0000-0003-1225-7084}\inst{\ref{aff14},\ref{aff15}}
\and R.~B.~Metcalf\orcid{0000-0003-3167-2574}\inst{\ref{aff13},\ref{aff14}}
\and M.~Schirmer\orcid{0000-0003-2568-9994}\inst{\ref{aff20}}
\and S.~Serjeant\orcid{0000-0002-0517-7943}\inst{\ref{aff21}}
\and C.~Tortora\orcid{0000-0001-7958-6531}\inst{\ref{aff22}}
\and M.~Vaccari\orcid{0000-0002-6748-0577}\inst{\ref{aff23},\ref{aff24},\ref{aff25}}
\and G.~Vernardos\orcid{0000-0001-8554-7248}\inst{\ref{aff26},\ref{aff27}}
\and M.~Walmsley\orcid{0000-0002-6408-4181}\inst{\ref{aff28},\ref{aff29}}
\and A.~Amara\inst{\ref{aff30}}
\and S.~Andreon\orcid{0000-0002-2041-8784}\inst{\ref{aff31}}
\and N.~Auricchio\orcid{0000-0003-4444-8651}\inst{\ref{aff14}}
\and H.~Aussel\orcid{0000-0002-1371-5705}\inst{\ref{aff32}}
\and C.~Baccigalupi\orcid{0000-0002-8211-1630}\inst{\ref{aff33},\ref{aff34},\ref{aff35},\ref{aff36}}
\and M.~Baldi\orcid{0000-0003-4145-1943}\inst{\ref{aff37},\ref{aff14},\ref{aff15}}
\and A.~Balestra\orcid{0000-0002-6967-261X}\inst{\ref{aff38}}
\and S.~Bardelli\orcid{0000-0002-8900-0298}\inst{\ref{aff14}}
\and A.~Basset\inst{\ref{aff39}}
\and P.~Battaglia\orcid{0000-0002-7337-5909}\inst{\ref{aff14}}
\and R.~Bender\orcid{0000-0001-7179-0626}\inst{\ref{aff18},\ref{aff17}}
\and D.~Bonino\orcid{0000-0002-3336-9977}\inst{\ref{aff40}}
\and E.~Branchini\orcid{0000-0002-0808-6908}\inst{\ref{aff41},\ref{aff42},\ref{aff31}}
\and M.~Brescia\orcid{0000-0001-9506-5680}\inst{\ref{aff43},\ref{aff22},\ref{aff44}}
\and J.~Brinchmann\orcid{0000-0003-4359-8797}\inst{\ref{aff45},\ref{aff46}}
\and A.~Caillat\inst{\ref{aff9}}
\and S.~Camera\orcid{0000-0003-3399-3574}\inst{\ref{aff47},\ref{aff48},\ref{aff40}}
\and V.~Capobianco\orcid{0000-0002-3309-7692}\inst{\ref{aff40}}
\and C.~Carbone\orcid{0000-0003-0125-3563}\inst{\ref{aff49}}
\and J.~Carretero\orcid{0000-0002-3130-0204}\inst{\ref{aff50},\ref{aff51}}
\and S.~Casas\orcid{0000-0002-4751-5138}\inst{\ref{aff52},\ref{aff2}}
\and F.~J.~Castander\orcid{0000-0001-7316-4573}\inst{\ref{aff53},\ref{aff54}}
\and M.~Castellano\orcid{0000-0001-9875-8263}\inst{\ref{aff55}}
\and G.~Castignani\orcid{0000-0001-6831-0687}\inst{\ref{aff14}}
\and S.~Cavuoti\orcid{0000-0002-3787-4196}\inst{\ref{aff22},\ref{aff44}}
\and A.~Cimatti\inst{\ref{aff56}}
\and C.~Colodro-Conde\inst{\ref{aff57}}
\and G.~Congedo\orcid{0000-0003-2508-0046}\inst{\ref{aff58}}
\and C.~J.~Conselice\orcid{0000-0003-1949-7638}\inst{\ref{aff29}}
\and L.~Conversi\orcid{0000-0002-6710-8476}\inst{\ref{aff59},\ref{aff5}}
\and Y.~Copin\orcid{0000-0002-5317-7518}\inst{\ref{aff60}}
\and L.~Corcione\orcid{0000-0002-6497-5881}\inst{\ref{aff40}}
\and F.~Courbin\orcid{0000-0003-0758-6510}\inst{\ref{aff6},\ref{aff61},\ref{aff62}}
\and H.~M.~Courtois\orcid{0000-0003-0509-1776}\inst{\ref{aff63}}
\and M.~Cropper\orcid{0000-0003-4571-9468}\inst{\ref{aff64}}
\and A.~Da~Silva\orcid{0000-0002-6385-1609}\inst{\ref{aff65},\ref{aff66}}
\and H.~Degaudenzi\orcid{0000-0002-5887-6799}\inst{\ref{aff67}}
\and G.~De~Lucia\orcid{0000-0002-6220-9104}\inst{\ref{aff34}}
\and A.~M.~Di~Giorgio\orcid{0000-0002-4767-2360}\inst{\ref{aff68}}
\and J.~Dinis\orcid{0000-0001-5075-1601}\inst{\ref{aff65},\ref{aff66}}
\and F.~Dubath\orcid{0000-0002-6533-2810}\inst{\ref{aff67}}
\and C.~A.~J.~Duncan\inst{\ref{aff29}}
\and X.~Dupac\inst{\ref{aff5}}
\and S.~Dusini\orcid{0000-0002-1128-0664}\inst{\ref{aff69}}
\and M.~Farina\orcid{0000-0002-3089-7846}\inst{\ref{aff68}}
\and S.~Farrens\orcid{0000-0002-9594-9387}\inst{\ref{aff32}}
\and F.~Faustini\orcid{0000-0001-6274-5145}\inst{\ref{aff70},\ref{aff55}}
\and S.~Ferriol\inst{\ref{aff60}}
\and N.~Fourmanoit\orcid{0009-0005-6816-6925}\inst{\ref{aff71}}
\and M.~Frailis\orcid{0000-0002-7400-2135}\inst{\ref{aff34}}
\and E.~Franceschi\orcid{0000-0002-0585-6591}\inst{\ref{aff14}}
\and M.~Fumana\orcid{0000-0001-6787-5950}\inst{\ref{aff49}}
\and S.~Galeotta\orcid{0000-0002-3748-5115}\inst{\ref{aff34}}
\and W.~Gillard\orcid{0000-0003-4744-9748}\inst{\ref{aff71}}
\and B.~Gillis\orcid{0000-0002-4478-1270}\inst{\ref{aff58}}
\and C.~Giocoli\orcid{0000-0002-9590-7961}\inst{\ref{aff14},\ref{aff15}}
\and B.~R.~Granett\orcid{0000-0003-2694-9284}\inst{\ref{aff31}}
\and A.~Grazian\orcid{0000-0002-5688-0663}\inst{\ref{aff38}}
\and F.~Grupp\inst{\ref{aff18},\ref{aff17}}
\and L.~Guzzo\orcid{0000-0001-8264-5192}\inst{\ref{aff72},\ref{aff31}}
\and S.~V.~H.~Haugan\orcid{0000-0001-9648-7260}\inst{\ref{aff73}}
\and J.~Hoar\inst{\ref{aff5}}
\and H.~Hoekstra\orcid{0000-0002-0641-3231}\inst{\ref{aff74}}
\and W.~Holmes\inst{\ref{aff75}}
\and I.~Hook\orcid{0000-0002-2960-978X}\inst{\ref{aff76}}
\and F.~Hormuth\inst{\ref{aff77}}
\and A.~Hornstrup\orcid{0000-0002-3363-0936}\inst{\ref{aff78},\ref{aff79}}
\and P.~Hudelot\inst{\ref{aff10}}
\and K.~Jahnke\orcid{0000-0003-3804-2137}\inst{\ref{aff20}}
\and M.~Jhabvala\inst{\ref{aff80}}
\and B.~Joachimi\orcid{0000-0001-7494-1303}\inst{\ref{aff81}}
\and E.~Keih\"anen\orcid{0000-0003-1804-7715}\inst{\ref{aff82}}
\and S.~Kermiche\orcid{0000-0002-0302-5735}\inst{\ref{aff71}}
\and A.~Kiessling\orcid{0000-0002-2590-1273}\inst{\ref{aff75}}
\and M.~Kilbinger\orcid{0000-0001-9513-7138}\inst{\ref{aff32}}
\and R.~Kohley\inst{\ref{aff5}}
\and B.~Kubik\orcid{0009-0006-5823-4880}\inst{\ref{aff60}}
\and M.~K\"ummel\orcid{0000-0003-2791-2117}\inst{\ref{aff17}}
\and M.~Kunz\orcid{0000-0002-3052-7394}\inst{\ref{aff83}}
\and H.~Kurki-Suonio\orcid{0000-0002-4618-3063}\inst{\ref{aff84},\ref{aff85}}
\and O.~Lahav\orcid{0000-0002-1134-9035}\inst{\ref{aff81}}
\and R.~Laureijs\inst{\ref{aff86},\ref{aff87}}
\and D.~Le~Mignant\orcid{0000-0002-5339-5515}\inst{\ref{aff9}}
\and S.~Ligori\orcid{0000-0003-4172-4606}\inst{\ref{aff40}}
\and P.~B.~Lilje\orcid{0000-0003-4324-7794}\inst{\ref{aff73}}
\and V.~Lindholm\orcid{0000-0003-2317-5471}\inst{\ref{aff84},\ref{aff85}}
\and I.~Lloro\orcid{0000-0001-5966-1434}\inst{\ref{aff88}}
\and G.~Mainetti\orcid{0000-0003-2384-2377}\inst{\ref{aff89}}
\and E.~Maiorano\orcid{0000-0003-2593-4355}\inst{\ref{aff14}}
\and O.~Mansutti\orcid{0000-0001-5758-4658}\inst{\ref{aff34}}
\and O.~Marggraf\orcid{0000-0001-7242-3852}\inst{\ref{aff90}}
\and K.~Markovic\orcid{0000-0001-6764-073X}\inst{\ref{aff75}}
\and M.~Martinelli\orcid{0000-0002-6943-7732}\inst{\ref{aff55},\ref{aff91}}
\and N.~Martinet\orcid{0000-0003-2786-7790}\inst{\ref{aff9}}
\and F.~Marulli\orcid{0000-0002-8850-0303}\inst{\ref{aff13},\ref{aff14},\ref{aff15}}
\and R.~Massey\orcid{0000-0002-6085-3780}\inst{\ref{aff92}}
\and E.~Medinaceli\orcid{0000-0002-4040-7783}\inst{\ref{aff14}}
\and S.~Mei\orcid{0000-0002-2849-559X}\inst{\ref{aff93}}
\and M.~Melchior\inst{\ref{aff94}}
\and Y.~Mellier\inst{\ref{aff95},\ref{aff10}}
\and E.~Merlin\orcid{0000-0001-6870-8900}\inst{\ref{aff55}}
\and G.~Meylan\inst{\ref{aff6}}
\and M.~Moresco\orcid{0000-0002-7616-7136}\inst{\ref{aff13},\ref{aff14}}
\and L.~Moscardini\orcid{0000-0002-3473-6716}\inst{\ref{aff13},\ref{aff14},\ref{aff15}}
\and R.~Nakajima\inst{\ref{aff90}}
\and R.~C.~Nichol\orcid{0000-0003-0939-6518}\inst{\ref{aff30}}
\and S.-M.~Niemi\inst{\ref{aff86}}
\and J.~W.~Nightingale\orcid{0000-0002-8987-7401}\inst{\ref{aff96}}
\and C.~Padilla\orcid{0000-0001-7951-0166}\inst{\ref{aff97}}
\and S.~Paltani\orcid{0000-0002-8108-9179}\inst{\ref{aff67}}
\and F.~Pasian\orcid{0000-0002-4869-3227}\inst{\ref{aff34}}
\and K.~Pedersen\inst{\ref{aff98}}
\and W.~J.~Percival\orcid{0000-0002-0644-5727}\inst{\ref{aff99},\ref{aff100},\ref{aff101}}
\and V.~Pettorino\inst{\ref{aff86}}
\and S.~Pires\orcid{0000-0002-0249-2104}\inst{\ref{aff32}}
\and G.~Polenta\orcid{0000-0003-4067-9196}\inst{\ref{aff70}}
\and M.~Poncet\inst{\ref{aff39}}
\and L.~A.~Popa\inst{\ref{aff102}}
\and L.~Pozzetti\orcid{0000-0001-7085-0412}\inst{\ref{aff14}}
\and F.~Raison\orcid{0000-0002-7819-6918}\inst{\ref{aff18}}
\and R.~Rebolo\inst{\ref{aff57},\ref{aff103},\ref{aff104}}
\and A.~Renzi\orcid{0000-0001-9856-1970}\inst{\ref{aff105},\ref{aff69}}
\and J.~Rhodes\orcid{0000-0002-4485-8549}\inst{\ref{aff75}}
\and G.~Riccio\inst{\ref{aff22}}
\and H.-W.~Rix\orcid{0000-0003-4996-9069}\inst{\ref{aff20}}
\and E.~Romelli\orcid{0000-0003-3069-9222}\inst{\ref{aff34}}
\and M.~Roncarelli\orcid{0000-0001-9587-7822}\inst{\ref{aff14}}
\and E.~Rossetti\orcid{0000-0003-0238-4047}\inst{\ref{aff37}}
\and B.~Rusholme\orcid{0000-0001-7648-4142}\inst{\ref{aff106}}
\and R.~Saglia\orcid{0000-0003-0378-7032}\inst{\ref{aff17},\ref{aff18}}
\and Z.~Sakr\orcid{0000-0002-4823-3757}\inst{\ref{aff107},\ref{aff108},\ref{aff109}}
\and A.~G.~S\'anchez\orcid{0000-0003-1198-831X}\inst{\ref{aff18}}
\and D.~Sapone\orcid{0000-0001-7089-4503}\inst{\ref{aff110}}
\and B.~Sartoris\orcid{0000-0003-1337-5269}\inst{\ref{aff17},\ref{aff34}}
\and P.~Schneider\orcid{0000-0001-8561-2679}\inst{\ref{aff90}}
\and T.~Schrabback\orcid{0000-0002-6987-7834}\inst{\ref{aff111}}
\and A.~Secroun\orcid{0000-0003-0505-3710}\inst{\ref{aff71}}
\and G.~Seidel\orcid{0000-0003-2907-353X}\inst{\ref{aff20}}
\and S.~Serrano\orcid{0000-0002-0211-2861}\inst{\ref{aff54},\ref{aff112},\ref{aff53}}
\and C.~Sirignano\orcid{0000-0002-0995-7146}\inst{\ref{aff105},\ref{aff69}}
\and G.~Sirri\orcid{0000-0003-2626-2853}\inst{\ref{aff15}}
\and L.~Stanco\orcid{0000-0002-9706-5104}\inst{\ref{aff69}}
\and J.~Steinwagner\orcid{0000-0001-7443-1047}\inst{\ref{aff18}}
\and P.~Tallada-Cresp\'{i}\orcid{0000-0002-1336-8328}\inst{\ref{aff50},\ref{aff51}}
\and I.~Tereno\inst{\ref{aff65},\ref{aff113}}
\and R.~Toledo-Moreo\orcid{0000-0002-2997-4859}\inst{\ref{aff114}}
\and F.~Torradeflot\orcid{0000-0003-1160-1517}\inst{\ref{aff51},\ref{aff50}}
\and I.~Tutusaus\orcid{0000-0002-3199-0399}\inst{\ref{aff108}}
\and L.~Valenziano\orcid{0000-0002-1170-0104}\inst{\ref{aff14},\ref{aff115}}
\and T.~Vassallo\orcid{0000-0001-6512-6358}\inst{\ref{aff17},\ref{aff34}}
\and G.~Verdoes~Kleijn\orcid{0000-0001-5803-2580}\inst{\ref{aff87}}
\and A.~Veropalumbo\orcid{0000-0003-2387-1194}\inst{\ref{aff31},\ref{aff42},\ref{aff116}}
\and Y.~Wang\orcid{0000-0002-4749-2984}\inst{\ref{aff117}}
\and J.~Weller\orcid{0000-0002-8282-2010}\inst{\ref{aff17},\ref{aff18}}
\and A.~Zacchei\orcid{0000-0003-0396-1192}\inst{\ref{aff34},\ref{aff33}}
\and G.~Zamorani\orcid{0000-0002-2318-301X}\inst{\ref{aff14}}
\and E.~Zucca\orcid{0000-0002-5845-8132}\inst{\ref{aff14}}
\and C.~Burigana\orcid{0000-0002-3005-5796}\inst{\ref{aff25},\ref{aff115}}
\and P.~Casenove\inst{\ref{aff39}}
\and A.~Mora\orcid{0000-0002-1922-8529}\inst{\ref{aff118}}
\and V.~Scottez\inst{\ref{aff95},\ref{aff119}}
\and M.~Viel\orcid{0000-0002-2642-5707}\inst{\ref{aff33},\ref{aff34},\ref{aff36},\ref{aff35},\ref{aff120}}
\and M.~Jauzac\orcid{0000-0003-1974-8732}\inst{\ref{aff121},\ref{aff92},\ref{aff122},\ref{aff123}}
\and H.~Dannerbauer\orcid{0000-0001-7147-3575}\inst{\ref{aff124}}}
										   
\institute{Max-Planck-Institut f\"ur Astrophysik, Karl-Schwarzschild-Str.~1, 85748 Garching, Germany\label{aff1}
\and
Institute of Cosmology and Gravitation, University of Portsmouth, Portsmouth PO1 3FX, UK\label{aff2}
\and
STAR Institute, Quartier Agora - All\'ee du six Ao\^ut, 19c B-4000 Li\`ege, Belgium\label{aff3}
\and
Sterrenkundig Observatorium, Universiteit Gent, Krijgslaan 281 S9, 9000 Gent, Belgium\label{aff4}
\and
ESAC/ESA, Camino Bajo del Castillo, s/n., Urb. Villafranca del Castillo, 28692 Villanueva de la Ca\~nada, Madrid, Spain\label{aff5}
\and
Institute of Physics, Laboratory of Astrophysics, Ecole Polytechnique F\'ed\'erale de Lausanne (EPFL), Observatoire de Sauverny, 1290 Versoix, Switzerland\label{aff6}
\and
SCITAS, Ecole Polytechnique F\'ed\'erale de Lausanne (EPFL), 1015 Lausanne, Switzerland\label{aff7}
\and
Department of Physics and Astronomy, University of California, Davis, CA 95616, USA\label{aff8}
\and
Aix-Marseille Universit\'e, CNRS, CNES, LAM, Marseille, France\label{aff9}
\and
Institut d'Astrophysique de Paris, UMR 7095, CNRS, and Sorbonne Universit\'e, 98 bis boulevard Arago, 75014 Paris, France\label{aff10}
\and
Technical University of Munich, TUM School of Natural Sciences, Physics Department, James-Franck-Str.~1, 85748 Garching, Germany\label{aff11}
\and
Department of Physics \& Astronomy, University of California Irvine, Irvine CA 92697, USA\label{aff12}
\and
Dipartimento di Fisica e Astronomia "Augusto Righi" - Alma Mater Studiorum Universit\`a di Bologna, via Piero Gobetti 93/2, 40129 Bologna, Italy\label{aff13}
\and
INAF-Osservatorio di Astrofisica e Scienza dello Spazio di Bologna, Via Piero Gobetti 93/3, 40129 Bologna, Italy\label{aff14}
\and
INFN-Sezione di Bologna, Viale Berti Pichat 6/2, 40127 Bologna, Italy\label{aff15}
\and
Instituto de F\'isica de Cantabria, Edificio Juan Jord\'a, Avenida de los Castros, 39005 Santander, Spain\label{aff16}
\and
Universit\"ats-Sternwarte M\"unchen, Fakult\"at f\"ur Physik, Ludwig-Maximilians-Universit\"at M\"unchen, Scheinerstrasse 1, 81679 M\"unchen, Germany\label{aff17}
\and
Max Planck Institute for Extraterrestrial Physics, Giessenbachstr. 1, 85748 Garching, Germany\label{aff18}
\and
FRACTAL S.L.N.E., calle Tulip\'an 2, Portal 13 1A, 28231, Las Rozas de Madrid, Spain\label{aff19}
\and
Max-Planck-Institut f\"ur Astronomie, K\"onigstuhl 17, 69117 Heidelberg, Germany\label{aff20}
\and
School of Physical Sciences, The Open University, Milton Keynes, MK7 6AA, UK\label{aff21}
\and
INAF-Osservatorio Astronomico di Capodimonte, Via Moiariello 16, 80131 Napoli, Italy\label{aff22}
\and
Inter-University Institute for Data Intensive Astronomy, Department of Astronomy, University of Cape Town, 7701 Rondebosch, Cape Town, South Africa\label{aff23}
\and
Inter-University Institute for Data Intensive Astronomy, Department of Physics and Astronomy, University of the Western Cape, 7535 Bellville, Cape Town, South Africa\label{aff24}
\and
INAF, Istituto di Radioastronomia, Via Piero Gobetti 101, 40129 Bologna, Italy\label{aff25}
\and
Department of Physics and Astronomy, Lehman College of the CUNY, Bronx, NY 10468, USA\label{aff26}
\and
American Museum of Natural History, Department of Astrophysics, New York, NY 10024, USA\label{aff27}
\and
David A. Dunlap Department of Astronomy \& Astrophysics, University of Toronto, 50 St George Street, Toronto, Ontario M5S 3H4, Canada\label{aff28}
\and
Jodrell Bank Centre for Astrophysics, Department of Physics and Astronomy, University of Manchester, Oxford Road, Manchester M13 9PL, UK\label{aff29}
\and
School of Mathematics and Physics, University of Surrey, Guildford, Surrey, GU2 7XH, UK\label{aff30}
\and
INAF-Osservatorio Astronomico di Brera, Via Brera 28, 20122 Milano, Italy\label{aff31}
\and
Universit\'e Paris-Saclay, Universit\'e Paris Cit\'e, CEA, CNRS, AIM, 91191, Gif-sur-Yvette, France\label{aff32}
\and
IFPU, Institute for Fundamental Physics of the Universe, via Beirut 2, 34151 Trieste, Italy\label{aff33}
\and
INAF-Osservatorio Astronomico di Trieste, Via G. B. Tiepolo 11, 34143 Trieste, Italy\label{aff34}
\and
INFN, Sezione di Trieste, Via Valerio 2, 34127 Trieste TS, Italy\label{aff35}
\and
SISSA, International School for Advanced Studies, Via Bonomea 265, 34136 Trieste TS, Italy\label{aff36}
\and
Dipartimento di Fisica e Astronomia, Universit\`a di Bologna, Via Gobetti 93/2, 40129 Bologna, Italy\label{aff37}
\and
INAF-Osservatorio Astronomico di Padova, Via dell'Osservatorio 5, 35122 Padova, Italy\label{aff38}
\and
Centre National d'Etudes Spatiales -- Centre spatial de Toulouse, 18 avenue Edouard Belin, 31401 Toulouse Cedex 9, France\label{aff39}
\and
INAF-Osservatorio Astrofisico di Torino, Via Osservatorio 20, 10025 Pino Torinese (TO), Italy\label{aff40}
\and
Dipartimento di Fisica, Universit\`a di Genova, Via Dodecaneso 33, 16146, Genova, Italy\label{aff41}
\and
INFN-Sezione di Genova, Via Dodecaneso 33, 16146, Genova, Italy\label{aff42}
\and
Department of Physics "E. Pancini", University Federico II, Via Cinthia 6, 80126, Napoli, Italy\label{aff43}
\and
INFN section of Naples, Via Cinthia 6, 80126, Napoli, Italy\label{aff44}
\and
Instituto de Astrof\'isica e Ci\^encias do Espa\c{c}o, Universidade do Porto, CAUP, Rua das Estrelas, PT4150-762 Porto, Portugal\label{aff45}
\and
Faculdade de Ci\^encias da Universidade do Porto, Rua do Campo de Alegre, 4150-007 Porto, Portugal\label{aff46}
\and
Dipartimento di Fisica, Universit\`a degli Studi di Torino, Via P. Giuria 1, 10125 Torino, Italy\label{aff47}
\and
INFN-Sezione di Torino, Via P. Giuria 1, 10125 Torino, Italy\label{aff48}
\and
INAF-IASF Milano, Via Alfonso Corti 12, 20133 Milano, Italy\label{aff49}
\and
Centro de Investigaciones Energ\'eticas, Medioambientales y Tecnol\'ogicas (CIEMAT), Avenida Complutense 40, 28040 Madrid, Spain\label{aff50}
\and
Port d'Informaci\'{o} Cient\'{i}fica, Campus UAB, C. Albareda s/n, 08193 Bellaterra (Barcelona), Spain\label{aff51}
\and
Institute for Theoretical Particle Physics and Cosmology (TTK), RWTH Aachen University, 52056 Aachen, Germany\label{aff52}
\and
Institute of Space Sciences (ICE, CSIC), Campus UAB, Carrer de Can Magrans, s/n, 08193 Barcelona, Spain\label{aff53}
\and
Institut d'Estudis Espacials de Catalunya (IEEC),  Edifici RDIT, Campus UPC, 08860 Castelldefels, Barcelona, Spain\label{aff54}
\and
INAF-Osservatorio Astronomico di Roma, Via Frascati 33, 00078 Monteporzio Catone, Italy\label{aff55}
\and
Dipartimento di Fisica e Astronomia "Augusto Righi" - Alma Mater Studiorum Universit\`a di Bologna, Viale Berti Pichat 6/2, 40127 Bologna, Italy\label{aff56}
\and
Instituto de Astrof\'isica de Canarias, Calle V\'ia L\'actea s/n, 38204, San Crist\'obal de La Laguna, Tenerife, Spain\label{aff57}
\and
Institute for Astronomy, University of Edinburgh, Royal Observatory, Blackford Hill, Edinburgh EH9 3HJ, UK\label{aff58}
\and
European Space Agency/ESRIN, Largo Galileo Galilei 1, 00044 Frascati, Roma, Italy\label{aff59}
\and
Universit\'e Claude Bernard Lyon 1, CNRS/IN2P3, IP2I Lyon, UMR 5822, Villeurbanne, F-69100, France\label{aff60}
\and
Institut de Ci\`{e}ncies del Cosmos (ICCUB), Universitat de Barcelona (IEEC-UB), Mart\'{i} i Franqu\`{e}s 1, 08028 Barcelona, Spain\label{aff61}
\and
Instituci\'o Catalana de Recerca i Estudis Avan\c{c}ats (ICREA), Passeig de Llu\'{\i}s Companys 23, 08010 Barcelona, Spain\label{aff62}
\and
UCB Lyon 1, CNRS/IN2P3, IUF, IP2I Lyon, 4 rue Enrico Fermi, 69622 Villeurbanne, France\label{aff63}
\and
Mullard Space Science Laboratory, University College London, Holmbury St Mary, Dorking, Surrey RH5 6NT, UK\label{aff64}
\and
Departamento de F\'isica, Faculdade de Ci\^encias, Universidade de Lisboa, Edif\'icio C8, Campo Grande, PT1749-016 Lisboa, Portugal\label{aff65}
\and
Instituto de Astrof\'isica e Ci\^encias do Espa\c{c}o, Faculdade de Ci\^encias, Universidade de Lisboa, Campo Grande, 1749-016 Lisboa, Portugal\label{aff66}
\and
Department of Astronomy, University of Geneva, ch. d'Ecogia 16, 1290 Versoix, Switzerland\label{aff67}
\and
INAF-Istituto di Astrofisica e Planetologia Spaziali, via del Fosso del Cavaliere, 100, 00100 Roma, Italy\label{aff68}
\and
INFN-Padova, Via Marzolo 8, 35131 Padova, Italy\label{aff69}
\and
Space Science Data Center, Italian Space Agency, via del Politecnico snc, 00133 Roma, Italy\label{aff70}
\and
Aix-Marseille Universit\'e, CNRS/IN2P3, CPPM, Marseille, France\label{aff71}
\and
Dipartimento di Fisica "Aldo Pontremoli", Universit\`a degli Studi di Milano, Via Celoria 16, 20133 Milano, Italy\label{aff72}
\and
Institute of Theoretical Astrophysics, University of Oslo, P.O. Box 1029 Blindern, 0315 Oslo, Norway\label{aff73}
\and
Leiden Observatory, Leiden University, Einsteinweg 55, 2333 CC Leiden, The Netherlands\label{aff74}
\and
Jet Propulsion Laboratory, California Institute of Technology, 4800 Oak Grove Drive, Pasadena, CA, 91109, USA\label{aff75}
\and
Department of Physics, Lancaster University, Lancaster, LA1 4YB, UK\label{aff76}
\and
Felix Hormuth Engineering, Goethestr. 17, 69181 Leimen, Germany\label{aff77}
\and
Technical University of Denmark, Elektrovej 327, 2800 Kgs. Lyngby, Denmark\label{aff78}
\and
Cosmic Dawn Center (DAWN), Denmark\label{aff79}
\and
NASA Goddard Space Flight Center, Greenbelt, MD 20771, USA\label{aff80}
\and
Department of Physics and Astronomy, University College London, Gower Street, London WC1E 6BT, UK\label{aff81}
\and
Department of Physics and Helsinki Institute of Physics, Gustaf H\"allstr\"omin katu 2, 00014 University of Helsinki, Finland\label{aff82}
\and
Universit\'e de Gen\`eve, D\'epartement de Physique Th\'eorique and Centre for Astroparticle Physics, 24 quai Ernest-Ansermet, CH-1211 Gen\`eve 4, Switzerland\label{aff83}
\and
Department of Physics, P.O. Box 64, 00014 University of Helsinki, Finland\label{aff84}
\and
Helsinki Institute of Physics, Gustaf H{\"a}llstr{\"o}min katu 2, University of Helsinki, Helsinki, Finland\label{aff85}
\and
European Space Agency/ESTEC, Keplerlaan 1, 2201 AZ Noordwijk, The Netherlands\label{aff86}
\and
Kapteyn Astronomical Institute, University of Groningen, PO Box 800, 9700 AV Groningen, The Netherlands\label{aff87}
\and
NOVA optical infrared instrumentation group at ASTRON, Oude Hoogeveensedijk 4, 7991PD, Dwingeloo, The Netherlands\label{aff88}
\and
Centre de Calcul de l'IN2P3/CNRS, 21 avenue Pierre de Coubertin 69627 Villeurbanne Cedex, France\label{aff89}
\and
Universit\"at Bonn, Argelander-Institut f\"ur Astronomie, Auf dem H\"ugel 71, 53121 Bonn, Germany\label{aff90}
\and
INFN-Sezione di Roma, Piazzale Aldo Moro, 2 - c/o Dipartimento di Fisica, Edificio G. Marconi, 00185 Roma, Italy\label{aff91}
\and
Department of Physics, Institute for Computational Cosmology, Durham University, South Road, DH1 3LE, UK\label{aff92}
\and
Universit\'e Paris Cit\'e, CNRS, Astroparticule et Cosmologie, 75013 Paris, France\label{aff93}
\and
University of Applied Sciences and Arts of Northwestern Switzerland, School of Engineering, 5210 Windisch, Switzerland\label{aff94}
\and
Institut d'Astrophysique de Paris, 98bis Boulevard Arago, 75014, Paris, France\label{aff95}
\and
School of Mathematics, Statistics and Physics, Newcastle University, Herschel Building, Newcastle-upon-Tyne, NE1 7RU, UK\label{aff96}
\and
Institut de F\'{i}sica d'Altes Energies (IFAE), The Barcelona Institute of Science and Technology, Campus UAB, 08193 Bellaterra (Barcelona), Spain\label{aff97}
\and
DARK, Niels Bohr Institute, University of Copenhagen, Jagtvej 155, 2200 Copenhagen, Denmark\label{aff98}
\and
Waterloo Centre for Astrophysics, University of Waterloo, Waterloo, Ontario N2L 3G1, Canada\label{aff99}
\and
Department of Physics and Astronomy, University of Waterloo, Waterloo, Ontario N2L 3G1, Canada\label{aff100}
\and
Perimeter Institute for Theoretical Physics, Waterloo, Ontario N2L 2Y5, Canada\label{aff101}
\and
Institute of Space Science, Str. Atomistilor, nr. 409 M\u{a}gurele, Ilfov, 077125, Romania\label{aff102}
\and
Departamento de Astrof\'isica, Universidad de La Laguna, 38206, La Laguna, Tenerife, Spain\label{aff103}
\and
Consejo Superior de Investigaciones Cientificas, Calle Serrano 117, 28006 Madrid, Spain\label{aff104}
\and
Dipartimento di Fisica e Astronomia "G. Galilei", Universit\`a di Padova, Via Marzolo 8, 35131 Padova, Italy\label{aff105}
\and
Caltech/IPAC, 1200 E. California Blvd., Pasadena, CA 91125, USA\label{aff106}
\and
Institut f\"ur Theoretische Physik, University of Heidelberg, Philosophenweg 16, 69120 Heidelberg, Germany\label{aff107}
\and
Institut de Recherche en Astrophysique et Plan\'etologie (IRAP), Universit\'e de Toulouse, CNRS, UPS, CNES, 14 Av. Edouard Belin, 31400 Toulouse, France\label{aff108}
\and
Universit\'e St Joseph; Faculty of Sciences, Beirut, Lebanon\label{aff109}
\and
Departamento de F\'isica, FCFM, Universidad de Chile, Blanco Encalada 2008, Santiago, Chile\label{aff110}
\and
Universit\"at Innsbruck, Institut f\"ur Astro- und Teilchenphysik, Technikerstr. 25/8, 6020 Innsbruck, Austria\label{aff111}
\and
Satlantis, University Science Park, Sede Bld 48940, Leioa-Bilbao, Spain\label{aff112}
\and
Instituto de Astrof\'isica e Ci\^encias do Espa\c{c}o, Faculdade de Ci\^encias, Universidade de Lisboa, Tapada da Ajuda, 1349-018 Lisboa, Portugal\label{aff113}
\and
Universidad Polit\'ecnica de Cartagena, Departamento de Electr\'onica y Tecnolog\'ia de Computadoras,  Plaza del Hospital 1, 30202 Cartagena, Spain\label{aff114}
\and
INFN-Bologna, Via Irnerio 46, 40126 Bologna, Italy\label{aff115}
\and
Dipartimento di Fisica, Universit\`a degli studi di Genova, and INFN-Sezione di Genova, via Dodecaneso 33, 16146, Genova, Italy\label{aff116}
\and
Infrared Processing and Analysis Center, California Institute of Technology, Pasadena, CA 91125, USA\label{aff117}
\and
Aurora Technology for European Space Agency (ESA), Camino bajo del Castillo, s/n, Urbanizacion Villafranca del Castillo, Villanueva de la Ca\~nada, 28692 Madrid, Spain\label{aff118}
\and
ICL, Junia, Universit\'e Catholique de Lille, LITL, 59000 Lille, France\label{aff119}
\and
ICSC - Centro Nazionale di Ricerca in High Performance Computing, Big Data e Quantum Computing, Via Magnanelli 2, Bologna, Italy\label{aff120}
\and
Department of Physics, Centre for Extragalactic Astronomy, Durham University, South Road, DH1 3LE, UK\label{aff121}
\and
Astrophysics Research Centre, University of KwaZulu-Natal, Westville Campus, Durban 4041, South Africa\label{aff122}
\and
School of Mathematics, Statistics \& Computer Science, University of KwaZulu-Natal, Westville Campus, Durban 4041, South Africa\label{aff123}
\and
Instituto de Astrof\'isica de Canarias (IAC); Departamento de Astrof\'isica, Universidad de La Laguna (ULL), 38200, La Laguna, Tenerife, Spain\label{aff124}}      

\abstract{
    We report the discovery of a complete Einstein ring around the elliptical galaxy \galaxyname, at $z=0.042$. This is the first strong gravitational lens discovered in \Euclid and the first in an NGC object from any survey. The combination of the low redshift of the lens galaxy, the brightness of the source galaxy ($\IE=18.1$ lensed, $\IE=21.3$ unlensed), and the completeness of the ring make this an exceptionally rare strong lens, unidentified until its observation by \Euclid. We present deep imaging data of the lens from the \Euclid Visible Camera (VIS) and Near-Infrared Spectrometer and Photometer (NISP) instruments, as well as resolved spectroscopy from the \textit{Keck} Cosmic Web Imager (KCWI). The \Euclid imaging in particular presents one of the highest signal-to-noise ratio optical/near-infrared observations of a strong gravitational lens to date. From the KCWI data we measure a source redshift of $z=0.406$. Using data from the Dark Energy Spectroscopic Instrument (DESI) we measure a velocity dispersion for the lens galaxy of $\sigma_\star=303\pm15\,\kms$. We model the lens galaxy light in detail, revealing angular structure that varies inside the Einstein ring. After subtracting this light model from the VIS observation, we model the strongly lensed images, finding an Einstein radius of $\ang{;;2.5}$, corresponding to $2.1\,\mathrm{kpc}$ at the redshift of the lens. This is small compared to the effective radius of the galaxy, $R_\mathrm{eff}\sim \ang{;;12.3}$. Combining the strong lensing measurements with analysis of the spectroscopic data we estimate a dark matter fraction inside the Einstein radius of $\dmfraction = (11.1_{-3.5}^{+5.4})\%$ and a stellar initial mass-function (IMF) mismatch parameter of $\imfmismatch = 1.26_{-0.08}^{+0.05}$, indicating a heavier-than-Chabrier IMF in the centre of the galaxy.
}


\titlerunning{An Einstein ring in NGC 6505}
\authorrunning{C. M. O'Riordan et al.}
   
\maketitle

\newpage
\section{\label{sec:introduction}Introduction}
In galaxy-galaxy strong gravitational lensing, the light from a distant source galaxy is distorted and magnified by the gravitational field of a foreground lens galaxy, such that multiple images of the source galaxy are formed. When the source is resolved, that is, not point-like, and close to the projected centre of the lens in the source plane a so-called `Einstein ring' is formed. Both Einstein rings and lensed point sources have tremendous scientific value and have been used in a wide variety of applications. When the source is time-varying, the time delay between the strongly lensed images can be used to constrain the Hubble constant
nt and other cosmological parameters \citep{Suyu2010,Wong2020,Birrer2020}. Precise modelling of the lensed images can be used to infer the properties of dark matter \citep[][and references in that review]{Minor2021,Bayer2023,Nightingale2024,Ballard2024,Vegetti2023}. Strong lenses can be used as `cosmic telescopes' to achieve higher spatial resolution when studying the lensed sources \citep{Barnacka2016,Oldham2017,Hartley2019}, and to test general relativity \citep{Collett2018}.

\begin{figure*}
	\centering
	\includegraphics[width=1.0\textwidth]{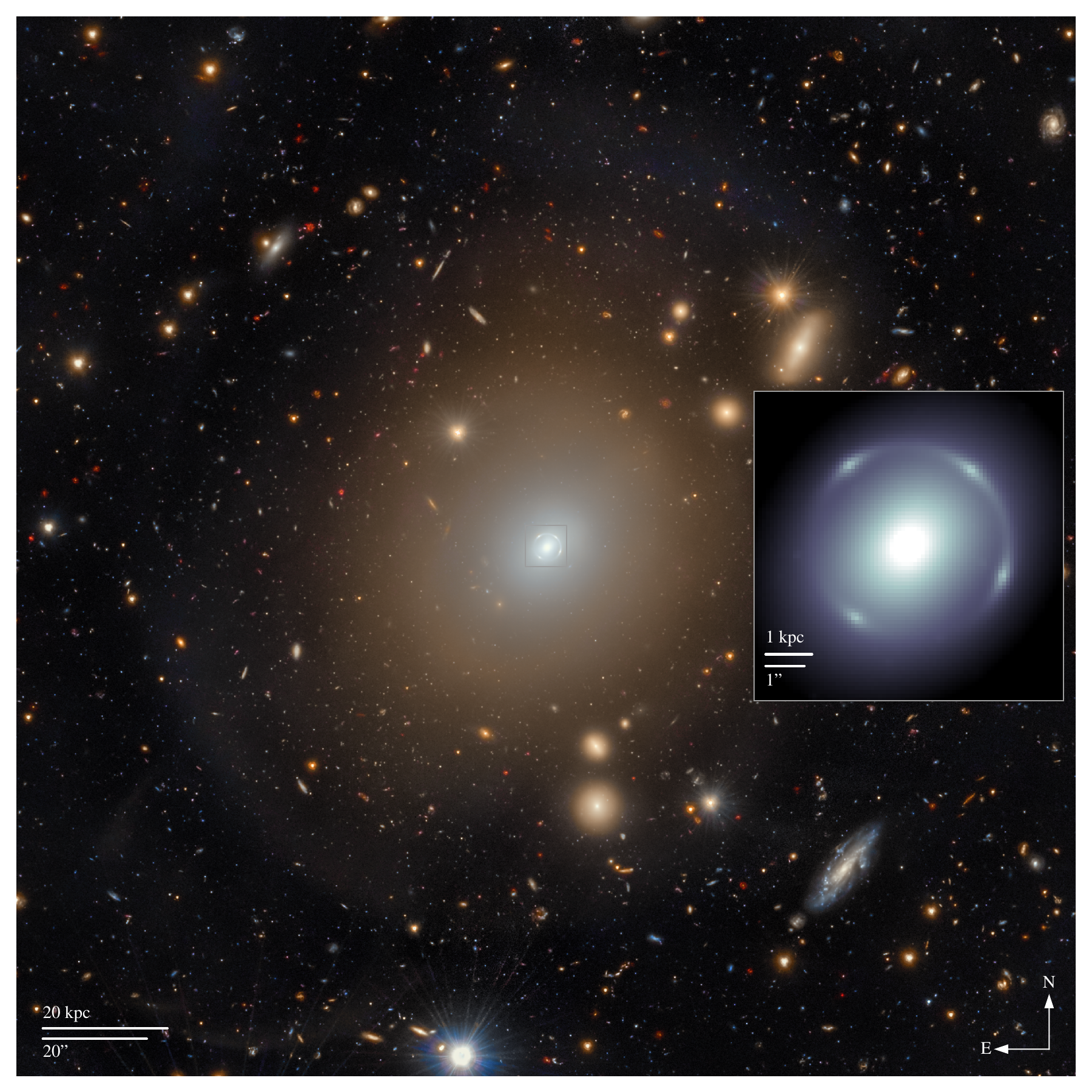}
	\caption{\Euclid imaging data used in this work, and in which \lensname~was discovered. The main panel shows a composite false colour image produced by combining the VIS and NISP data. The higher resolution broadband VIS \IE image is used to set the brightness, with the colour provided by the lower resolution NISP \YE, \JE, and \HE passband images. The central light of the galaxy is suppressed to make the lensed arc more visible. The inset shows only the higher resolution VIS data in the central $\ang{;;8}$ of the image, indicated by the square in the main panel. The angular scale and the physical scale at the redshift of the lens are given in each frame.}
	\label{fig:vis-image}
\end{figure*}

The most prevalent application of galaxy-scale strong lensing is in studying the lens itself, which is most often an early-type galaxy (ETG). Methods to separate the dark matter component from the stellar mass have not yet reached a consensus on the inner dark matter structure of these systems \citep{Sonnenfeld2012, Oldham2018, Shajib2021} but are building up a picture of ETGs as having bottom-heavy stellar initial mass functions (IMFs) in their inner regions \citep[e.g.][]{Auger2010} which decline to more Milky Way-like IMFs within the inner few kpc \citep{Collett2018, Sonnenfeld2018}. This view is supported by studies using dynamics \citep{Oldham2018b, Mehrgan2024} and stellar population modelling \citep{Conroy2017, Sarzi2018}, and seems to point towards a two-phase evolution scenario in which the inner regions of the ETG form at early times in situ and the outer envelope is accreted later, the two components being distinct and therefore having distinct stellar population properties \citep{Oser2010, Naab2014}. Low-redshift lenses are unique laboratories for testing this picture because the Einstein radius is typically at a smaller physical scale that encloses only the central region of the lens galaxy. The mass within the Einstein radius of these objects is dominated by stellar mass, thus facilitating a robust measure of the central stellar mass and IMF. So far, only $\sim 5$ such systems are known: J2237$+$0305 \citep{Huchra1985}, J1343$-$3810, J2100$-$4225, J0141$-$0735 \citep[SNELLS,][]{Smith2015}, and J0403$-$0239 \citep{Galbany2018}.

In this paper we present the discovery of a new low-redshift strong gravitational lens around NGC 6505 at $z=0.042$. The lens galaxy has been known since \citet{Swift1886}, and observed in \citet[][X-rays]{Henry1995} and \citet[][radio]{Brinkmann1999}. However, its Einstein ring was unknown until its observation by the \Euclid space telescope. The primary scientific goal of \Euclid is to obtain cosmological constraints from weak lensing and galaxy clustering over $14\,000$~deg$^2$ of the sky \citep[see][for a mission overview]{Euclid-Overview}. As a consequence of observing such a large area at the depth and resolution of \Euclid, it is forecast to discover $> 10^5$ new strong gravitational lenses \citep{Collett2015}. Results based on \Euclid Early Release Observations show encouraging evidence that this forecast is correct \citep{AcevedoBarroso2024}.

Here we present imaging data for this new strong lens from the \Euclid VIS and NISP instruments \citep[see][for detailed descriptions]{Euclid-VIS,Euclid-NISP}, as well as spectroscopy collected from the \emph{Keck} Cosmic Web Imager \citep[KCWI,][]{Morrissey2018}. We also make use of data from the Dark Energy Spectroscopic Instrument (DESI). From these data, we obtain results on the properties of the lens galaxy, and the redshift of the lensed source.

The paper is organised as follows. In Sect.~\ref{sec:data}, we describe the imaging and spectroscopic data used in the rest of the paper. In Sect.~\ref{sec:results}, we present modelling results for the lens galaxy and the strongly lensed emission. In Sect.~\ref{sec:discussion}, we discuss the rarity of the lens and its properties in the context of similar objects. In Sect.~\ref{sec:summary}, we summarise our results. Throughout the paper we assume a Planck 2015 cosmology with $H_0=67.8\,\kmsMpc$ and $\Omega_\mathrm{m}=0.308$ \citep{Planck2016}.

\section{\label{sec:data}Data}
\subsection{Euclid VIS and NISP imaging}

The lens was discovered serendipitously by B. Altieri when looking at early data in the Performance Verification (PV) phase of the `contamination scan' CALBLOCK-F-014 aimed at checking ice contamination in the Phase Diversity Campaign (PDC). Although the initial observation was deliberately defocussed as part of the PDC, the Einstein ring was clearly visible, and its nature was confirmed in the subsequent focussed observations. Following the discovery, we adopted the nickname `Altieri's lens' for this spectacular object. Later, \galaxyname~was observed in the first visit to the Euclid Deep Field North (EDF-N) at the edge of the field.

Figure~\ref{fig:vis-image} shows the imaging data from VIS \IE and NISP \YE, \JE, and \HE. Table~\ref{tab:data-properties} gives their properties. The data combine four observations from separate visits to the self calibration field, a subset of EDF-N, during the PV phase (Observation IDs: 65765, 65776, 65803, and 65811). Each observation includes tens of exposures. The result is much deeper imaging than in typical \Euclid Wide or Deep observing configurations. The individual VIS and NISP exposures are processed by their respective pipelines before being combined with external data from ground-based surveys into a single data product by the MERge processing pipeline (MER). The MER pipeline resamples all individual observations, with different pixel scales and pointings, on to a north-aligned grid with the pixel scale of VIS, that is, $\ang{;;0.1}$. The deep VIS and NISP data used here were processed by MER in August 2024, and are available in the MER tile with index $101832848$.

\begin{table}
	\centering
	\caption{Imaging data properties}
	\begin{tabular}{ p{2.15cm} r p{1.77cm} r r }
		Filter & Pixel scale & \raggedleft PSF FWHM & $N_\mathrm{exp}$ & $t_\mathrm{int}$\\
		& [arcsec] &  \raggedleft [arcsec] & & [s]\\
		\hline
		\Euclid/\IE & $0.100$ & \raggedleft $0.16$ & 122 & $40\,123$\\
		\Euclid/\YE,\JE,\HE &  $0.300$ & \raggedleft $0.35$ & $70$ & $6107$\\
		\hline
		CFIS/$u$ & $0.185$ & \raggedleft $\sim 0.70$ & - & $640$ \\
		CFIS/$r$ & $0.185$ & \raggedleft $\sim 0.70$ & - & $1664$ \\
		Pan-STARRS/$i$ & $0.258$ & \raggedleft $\sim 0.80$ & - & - \\
		HSC/$g$ & $0.170$ & \raggedleft $\sim 0.70$ & - & - \\
		HSC/$z$ & $0.170$ & \raggedleft $\sim 0.70$ & - & - \\
		\hline
	\end{tabular}
	\tablefoot{Columns are: the telescope and filter used, the image pixel scale, the full-width at half-maximum (FWHM) of the instrument point spread function (PSF), the total number of exposures, $N_\mathrm{exp}$, from each instrument at the location of \galaxyname, and the total integration time, $t_\mathrm{int}$, of those exposures. The PSF FWHM for the ground-based imaging is an estimate and in practice depends on observing conditions. Exposure information for some ground-based imaging was unavailable.}
	\label{tab:data-properties}
\end{table}

\subsection{Ancillary imaging}

Ground-based observations serve as supplementary imaging data to complement \Euclid's VIS and NISP imaging, and are especially used to ensure \Euclid obtains accurate photometric redshifts. In 2017, The Ultraviolet Near Infrared Optical Northern Survey (UNIONS) collaboration was established to provide additional data, particularly in optical and ultraviolet wavelengths, that extend the wavelength coverage of \Euclid \citep{Euclid-Overview}. The ancillary imaging consists of mosaic observations from three facilities; the Canada-France-Hawaii Telescope (CFHT)/Canada-France Imaging Survey (CFIS) \citep{Ibata_2017ApJ...848..128I}, the Panoramic Survey Telescope and Rapid Response System \citep[Pan-STARRS;][]{Chambers_2016arXiv161205560C}, and Subaru/Hyper Suprime-Cam (HSC) \citep{Miyazaki_2018PASJ...70S...1M}. In particular, the $u$ and $r$ bands are obtained using the CFHT, Pan-STARRS provides an image in the $i$ band, while observations in the $g$ and $z$ bands were collected with Subaru. The properties of these data are given in Table~\ref{tab:data-properties} with that of the \Euclid data. The UNIONS data used here are available in the MER tile with index $102158889$.

\begin{figure*}
	\centering
	\includegraphics[width=\textwidth]{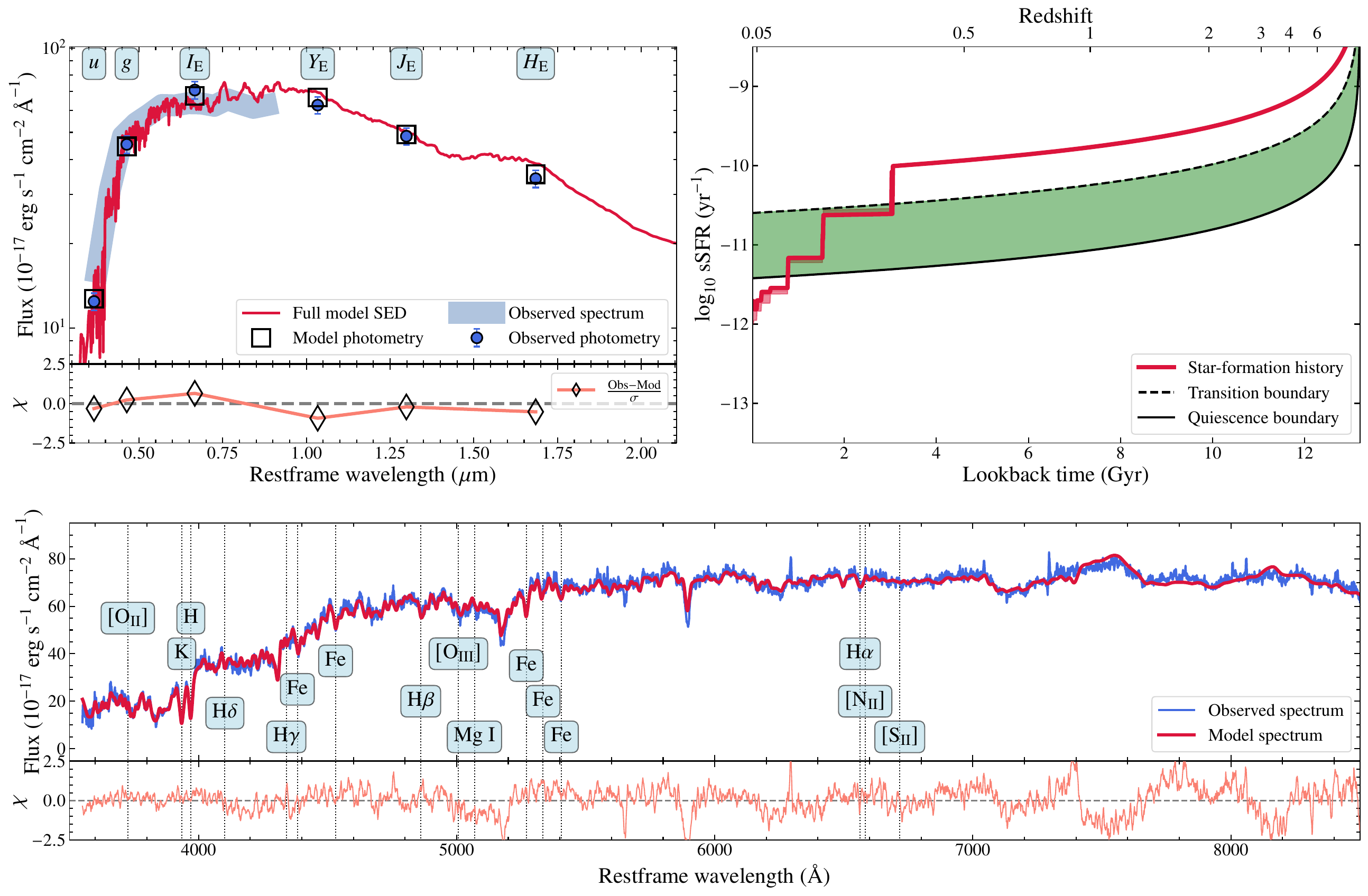}
	\caption{Photometry and spectroscopy for the lensing galaxy. \textit{Top left}: in the upper panel, blue points indicate the observed photometry from CFHT (MegaCam-$u$), Subaru (HSC-$g$), and \Euclid (\IE, \YE, \JE, and \HE), while open squares indicate the model photometry of the same bands. The red line shows the best-fit model SED, and the grey region marks the spectral region from DESI. The lower panel shows the residuals, defined by $\left(\mathrm{data} - \mathrm{model}\right)/\sigma$. \textit{Top right}: The recovered SFH (red line) and the transition boundary (green shaded area) from star forming (dashed black line) to quiescence (solid black line). \textit{Bottom}: in the upper panel, the spectroscopic data from DESI, with $\zl = 0.0424$, are shown in blue, while the best-fit model spectrum is shown in red. The vertical dotted lines indicate the central wavelength of various spectral features. The associated lower panel shows the residuals, defined by $\left(\mathrm{data} - \mathrm{model}\right)/\sigma$.}
	\label{fig:best_fit_SED_spectrum_and_SFH}
\end{figure*}

\begin{figure*}
	\centering
	\includegraphics[width=1.0\textwidth]{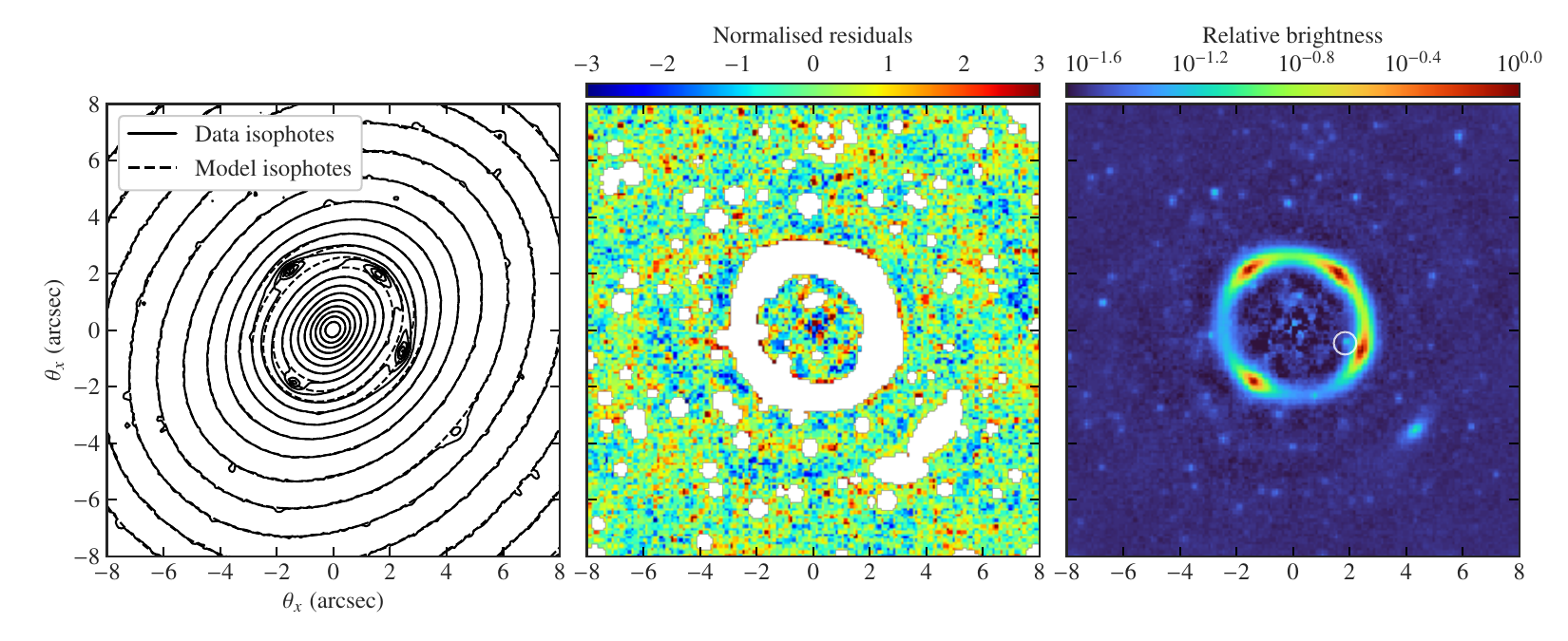}
	\caption{Modelling of the lens galaxy light profile. \textit{Left}: Isophotes from the final light model plotted with data isophotes in the inner region of NGC 6505. The lensed source emission and emission from other compact sources are visible in the data isophotes. \textit{Middle}: The difference between the data and the model, normalised by the noise in each pixel. Empty pixels are those masked from the fitting procedure. The residuals here show no evidence of a central strongly lensed image. \textit{Right}: The lens light subtracted data, to be used for strong lens modelling presented in Sect.~\ref{sec:lens-modelling}. The white circle indicates the compact source of emission included in the lens model.}
	\label{fig:vis-light-model}
\end{figure*}

\subsection{Keck Cosmic Web Imager and other spectroscopy}
\label{sec:kcwi-spectroscopy}

We obtained spectroscopy of the central $\ang{;;16.5}\times\ang{;;20.4}$ of \galaxyname~with the KCWI using the medium slicer and the low resolution grating ($R\sim$ 1800, rectangular pixels $\ang{;;0.29}\times\ang{;;0.68}$). We obtained a single $1000\,\mathrm{s}$ exposure on-target and a single $300\,\mathrm{s}$ exposure on-sky for the blue side (wavelength range $3500\,\AA$--$5600\,\AA$) and three $300\,\mathrm{s}$ exposures on-target and two $140\,\mathrm{s}$ exposures on-sky for the red side (wavelength range $5400\,\AA$--$10\,800\,\AA$) at twilight on the night of 31st March 2024. Both sides were reduced using the default settings of the KCWI Data Extraction and Reduction Pipeline \citep[{\tt kderp},][]{kderp2023}, and the datacubes were flux calibrated using observations of standard stars observed on the same night.\footnote{\url{https://github.com/Keck- DataReductionPipelines/KcwiDRP}} Kinematics measured from these spectra will be presented in a future paper.

We also retrieved the reduced spectrum of \galaxyname~from the DESI early data release.\footnote{\url{https://data.desi.lbl.gov/doc/releases/edr/}} The spectrum has a median signal-to-noise ratio of $\sim$140 and an aperture diameter of $\ang{;;1.5}$ (hence not including the source galaxy). 

\section{\label{sec:results}Results}

\subsection{Lens galaxy spectrum}
\label{sec:lens-kinematics}
\newcommand{\mstarChab}{M_{\star}^\mathrm{Chab}}

We first extracted kinematics from the DESI spectrum following the method described in \cite{Oldham2017c}, using stellar templates of A, F, G and K stars from the Indo-US Stellar Library of Coud\'{e} Feed Stellar Spectra \citep{Valdes2004} and modelling the spectrum as the sum of the light from these stars and a continuum component which we treated as a seventh-order polynomial. We tested the robustness of our extraction by repeating the modelling process with {\tt pPXF} \citep{ppxf} and found that both methods agreed within 1$\sigma$. We measure a central velocity dispersion $\sigma_\star=303\pm15\,\kms$ within the $\ang{;;1.5}$ DESI aperture, imposing a 5\% uncertainty floor due to the choice of stellar templates.

Following this, we used {\tt Prospector} \citep{Leja_2017ApJ...837..170L, Johnson_2021ApJS..254...22J} to simultaneously fit the photometric spectral energy distribution (SED) and DESI spectrum within an aperture of $\ang{;;1.5}$ in diameter. We use the broadband photometry of the MegaCam-$u$ from CFHT, the HSC-$g$ band from Subaru, and the four Euclid bands (\IE, \YE, \JE, and \HE). We do not include the MegaCam-$r$, and HSC-$z$ bands in the fitting due to the overlap of their filter response curve with the one of \IE band from \Euclid. Furthermore, the Pan-STARRS-$i$ flux is significantly lower than the rest of the SED, rendering it essentially unusable.  

{\tt Prospector} is an SED fitting code that utilises Bayesian and Monte Carlo sampling techniques to explore a multidimensional parameter space. {\tt Prospector} is capable of generating gridless, `on-the-fly' SEDs by integrating stellar, nebular, and dust models into composite stellar populations. One of the key strengths of {\tt Prospector} is its ability to handle photometric and spectroscopic data with a flexible spectrophotometric calibration. Additionally, it offers nonparametric star-formation histories (SFHs) with various prior distributions and parameterizations, allowing for the analysis of SFHs without imposing a specific functional shape.

We generated a physical model of 21 free parameters based on recent spectrophotometric studies \citep[e.g.][Nersesian et al. in prep]{Johnson_2021ApJS..254...22J, Tacchella_2022ApJ...926..134T}. We adopted the MILES stellar library \citep{Sanchez_Blazquez_2006MNRAS.371..703S} and the MIST isochrones \citep{Choi_2016ApJ...823..102C}, from the Flexible Stellar Populations Synthesis ({\tt FSPS}) code \citep{Conroy_2009ApJ...699..486C, Conroy_2010ApJ...712..833C}, and a \cite{Chabrier_2003PASP..115..763C} IMF. We used a `smooth' nonparametric SFH with a Student's-t prior distribution \citep{Leja_2019ApJ...876....3L} and ten time elements, based on the regularisation schemes by \citet{Ocvirk_2006MNRAS.365...46O} and \citet{Tojeiro_2007MNRAS.381.1252T}. Finally, we treated the dust effects in the UV and optical regime by using a flexible two-component attenuation law \citep{Charlot_2000ApJ...539..718C, Noll_2009A&A...507.1793N, Kriek_2013ApJ...775L..16K}, that differentiates the dust attenuation in birth-clouds from the diffuse dust in the interstellar medium (ISM). An error floor of 5\% is introduced to account for various systematics in the underlying stellar models \citep[e.g. the thermally pulsing asymptotic giant branch (TP-AGB) phase, convective overshooting, stellar remnants][]{Maraston_2006ApJ...652...85M, Conroy_2013ARA&A..51..393C}. 

The observational data of \galaxyname~from \Euclid, CFHT, and DESI along with the fitted model are shown in Fig.~\ref{fig:best_fit_SED_spectrum_and_SFH}. We measure a redshift for the galaxy of $\zl = 0.04243 \pm 0.00003$. It is evident that our model fits the observations well, with the residuals, for both photometry and spectroscopy, being distributed very close to 0. Another measure of the goodness of the fit is the $\chi^2$ statistic, defined as:

\begin{equation} \label{eq:reduced_chi2}
	\chi^2 = \sum_{i=1} \frac{\left(O_i - P_i\right)^2}{\sigma_i^2},
\end{equation}

\noindent where $O_i$ represent the flux densities of the observations, $P_i$ are the model fluxes, and $\sigma_i$ their uncertainties. The number of degrees of freedom, $N_\mathrm{dof}$, is calculated as the number of wavelength samples $N_\mathrm{\lambda,~spec}+N_\mathrm{\lambda,~phot}$ minus the free parameters in our model. We measure $\chi^2=42.3$ with $N_\mathrm{dof}=21$. 


From the best-fit model we measure the physical properties of the central $\ang{;;1.5}$ of \galaxyname. In particular, we measure a stellar mass of $M_{\star} = (2.51 \pm 0.06) \times 10^{10} M_{\odot}$ assuming a \citet{Chabrier_2003PASP..115..763C} IMF, a stellar metallicity of $\logten\left(Z_\star\right / \mathrm{Z}_\odot) = 0.181^{+0.006}_{-0.009}$, a mass-weighted stellar age of $t_\star = 9.01^{+0.22}_{-0.26}$~Gyr, and a SFR $= 0.01^{+0.003}_{-0.005}\,\Msun$~yr$^{-1}$.\footnote{The mass-weighted stellar age is the lookback time when 50\% of the stellar mass has been formed.} We also measure a stellar velocity dispersion of $\sigma_\star = 301 \pm 9$~km~s$^{-1}$, in excellent agreement with our {\tt pPXF} measurement. When necessary, to convert the stellar mass and SFR from \citet{Chabrier_2003PASP..115..763C} to \citet{Salpeter_1955ApJ...121..161S} IMFs, we can rescale these quantities by dividing them with a constant factor of 0.61 and 0.63, respectively.

From our analysis, we also recover the SFH of the very central $\ang{;;1.5}$ of NGC~6505. We define the SFH using the sSFR($t$), which is a better indicator of quiescence, defined as the fitted SFR of each time bin in our physical model, normalised by the total stellar mass formed up to that time bin. We also plot the transition boundary (green region) from star forming (${\rm sSFR} = 1 / \left[3~t_{\rm H}\right]$) to quiescence (sSFR$ = 1/\left[20~t_{\rm H}\right]$), as defined by \citet[][see also \citealp{Pacifici_2016ApJ...832...79P}]{Tacchella_2022ApJ...926..134T}, where $t_{\rm H}$ is the Hubble time at a redshift $z = 0.0424$. We find a gradually declining star formation with time up to redshift $\sim 0.5$. Then, at a lookback time of $\sim 3$~Gyr, the galaxy started slowly transitioning from actively star forming to quiescence, a period that lasted 2.3~Gyr.

\subsection{Lens galaxy light distribution}
\label{sec:lens-light-distribution}

We model the lens galaxy light profile in a square $\ang{;;8}\times\ang{;;8}$ cutout of the \IE image, centred on the brightest pixel of the lens galaxy, using the {\tt Galfit} code \citep{Peng2010}. We use a composite of seven S\'ersic profiles, each with independent multipole perturbations. The model fit is performed iteratively, adding complexity until the reduced chi-squared statistic, $\chisqnu=\chi^2/N_\mathrm{dof}$, does not improve. An iterative fit also allows us to progressively mask the many compact sources in the lens galaxy light, which are not visible until most of the lens light is removed. The lensed source emission is also masked. A detailed model of the light profile gives us lens light-subtracted data for strong lens modelling, as well as information about the structure of the galaxy, interior to the Einstein ring.

\begin{figure}
	\centering
	\includegraphics[width=1.0\columnwidth]{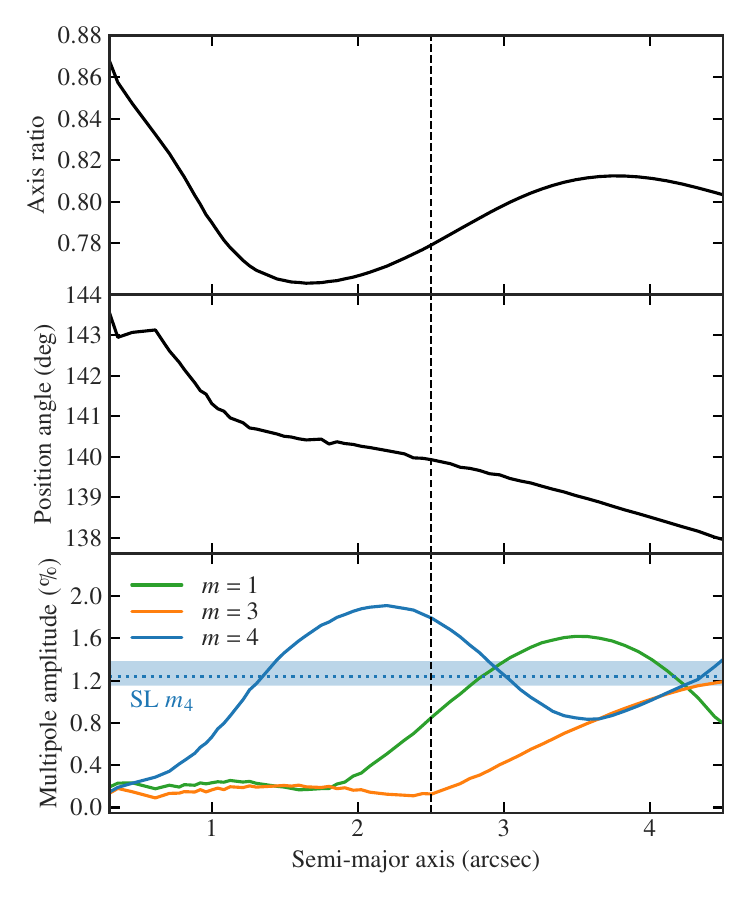}
	\caption{Properties of the lens galaxy isophotes as a function of the elliptical semi-major axis. \textit{Top panel}: the axis ratio, $q=b/a$, \textit{middle panel}: the position angle of the major axis, \textit{bottom panel}: the total amplitude of multipole perturbations of orders $m=1,3,4$. The multipole perturbations are formulated such that a positive order $m=4$ amplitude produces boxiness. The dotted horizontal line and shaded area indicate the median $m=4$ amplitude measured by the lens modelling only, and its uncertainty respectively. The vertical dashed line indicates the elliptical radius at the critical curve, equivalent to the Einstein radius.}
	\label{fig:isophote-properties}
\end{figure}

Figure~\ref{fig:vis-light-model} shows the light model fit and the resulting light-subtracted data used for strong lens modelling in Sect.~\ref{sec:lens-modelling}. The final model has $\chi^2=21474$ with $\ndof=21040$. In Fig.~\ref{fig:isophote-properties}, we plot the properties of the isophotes as a function of radius. Isophotes are first fit with ellipses and the position angle, $\posa$, and axis ratio, $q$, of the fitted ellipse are measured. Fourier modes, $m={1,3,4}$, are also fit to the isophotes to measure the strength of the non-elliptical components. If the elliptical radius at an angular position $(\theta_x,\theta_y)$ is defined $\etheta^2=(q\theta_x)^2+\theta_y^2$, then the multipole perturbations act as a small correction to the elliptical radius as a function of angle, $\phi$, such that
\begin{equation}
	\etheta'(\etheta,\phi) = \etheta + \sum_{m=\{1,3,4\}}\left[a_m\cos{(\phi-\posa)}+b_m\sin{(\phi-\posa)}\right],
\end{equation}
where $a_m$ and $b_m$ are the cosine and sine amplitudes of each multipole with order $m$. Whenever a `multipole amplitude' is discussed, this is the total strength of a single mode, or the quadratic sum of $a_m$ and $b_m$.
These modes are aligned with the position angle such that a positive mode $m=4$ indicates boxiness. The $m=2$ perturbation is not included as it describes simple ellipticity, already part of the model. No evidence is found for modes beyond $m=4$. A visual explanation of the effect of the different modes is available in \citet[][their Fig.~2]{ORiordan2024}. The multipole amplitudes can be interpreted as the distance that a perturbed isophote would be from its purely elliptical equivalent, relative to the size of the semi-major axis.

We find a positive ellipticity gradient between the centre of the galaxy and the Einstein radius, as well as a slight anti-clockwise isophote twist. We also find significant boxiness that peaks at an amplitude of $2\%$ just inside the Einstein radius. At larger radii there are also $m=1$ and $m=3$ perturbations. The perturbations found here are very typical of the isophotes of elliptical galaxies \citep{Hao2006}.

After subtracting the lens light model from the \IE image, the S/N in the brightest pixels of the lensed images exceeds $300$. Following the method of \citet{ORiordan2019}, we calculate the integrated S/N in a mask which includes all strongly lensed emission $2\sigma$ above the background. The integrated S/N with this definition is $1490$. This is extremely high for an optical strong lens observation.

Finally, we measure the stellar mass within the Einstein radius, found by strong lens modelling in Sec~\ref{sec:lens-modelling} to be $\ang{;;2.5}$. We repeat the light modelling with \texttt{Galfit} in the other photometric bands (\YE,\JE, and \HE) using the same composite of S\'ersic profiles. We then perform aperture photometry on the {\tt Galfit} model within the Einstein radius, and fit the photometric SED with {\tt Prospector}, finding $\mstarChab = (1.06 \pm 0.02) \times 10^{11}\Msun$.

\begin{figure}
	\centering
	\includegraphics[width=1.0\columnwidth]{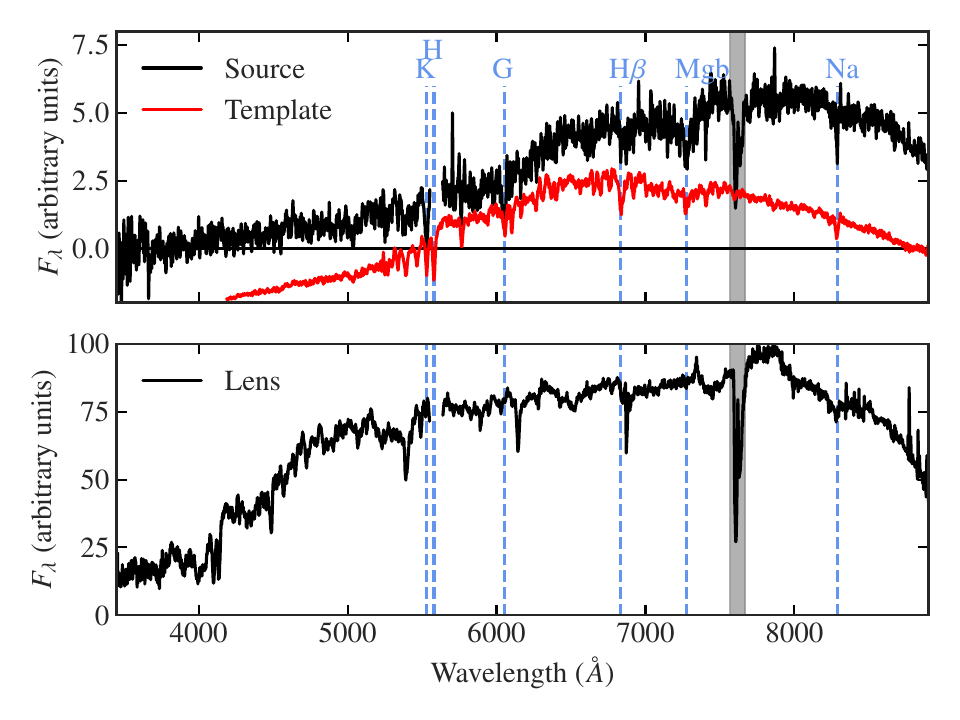}
	\caption{\textit{Upper}: KCWI spectrum of the lensed source galaxy smoothed with a 5 pixels kernel, and, \textit{lower}: of the lens galaxy. The shaded grey region marks sky absorption. The main absorption lines from the source are identified with blue dashed lines on both panels. A template of an ETG (offset vertically to ease legibility) at the same redshift as the source is also shown for comparison. Differences in the global shapes of the continua of the source and template reflect uncertainties in the absolute flux calibration of the KCWI data. The absence of coincidence between the absorption lines from the lens galaxy and from the source supports marginal contamination from the lens and confirms the robustness of the source redshift estimate. }
	\label{fig:spec-source}
\end{figure}

\begin{figure*}
	\centering
	\includegraphics[width=1.0\textwidth]{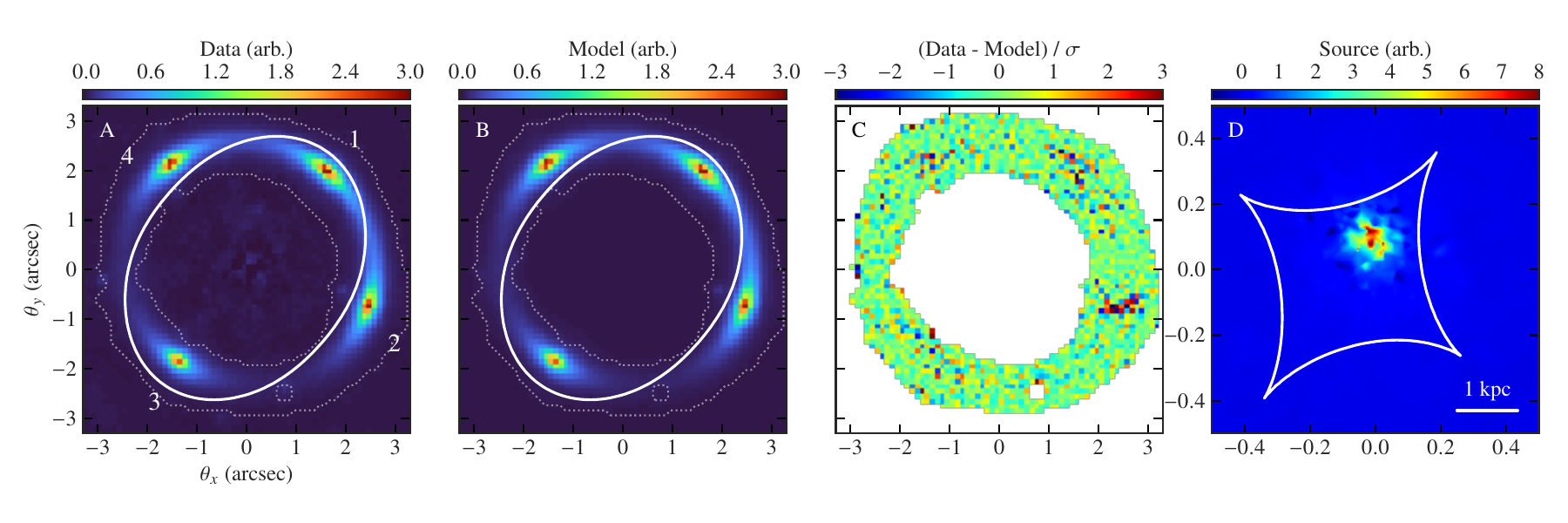}
	\caption{Elliptical power-law plus external shear and multipole model of the strongly lensed images. (A) the VIS data with the four lensed images labelled as in Table~\ref{tab:image-properties}, (B) the maximum a posteriori model, both in arbitrary flux units, (C) the normalised residuals, and (D) the pixellated source reconstruction in the same units as the data and model. Solid curves in the image plane and source plane are the critical curves and caustics respectively. In (A) and (B) the mask used to model the data is indicated by the dotted white shapes. The physical scale in the source plane is indicated.}
	\label{fig:lens-modelling}
\end{figure*}

\subsection{Source and lens redshift}
\label{sec:source-redshift}

The white-light images built by summing up the wavelengths from the red arm of the KCWI data clearly display an excess of flux at the location of the lensed images identified in \Euclid data. The contribution of the lens galaxy dominates over the source flux, preventing a clear spectral deblending of the lens and source from those raw data. Consequently, we subtract a 2D luminosity profile of the lens galaxy for each wavelength slice. We could not use the \Euclid-derived luminosity profile of the lens as a model because of the absence of a point-like image in the KCWI data. Instead, we use a single S\'ersic profile, with parameters initialised based on the observed \Euclid VIS morphology. Following a procedure similar to \citet{Braibant2014} and \citet{Sluse2019}, this model is fit to each wavelength slice using a Levenberg--Marquardt algorithm, which minimises a $\chi^2$ merit function. An additional pedestal flux, constant over the frame but different at each wavelength, has been added to that model. Since the lensed image flux contributes between $4\%$ and $20\%$ of the total flux (at $\sim$ 6000\,\AA), even an approximate model of the lens light, that is, a single S\'ersic, enables us to isolate the source with reasonable accuracy. The same procedure has been applied to the red and blue arms of the KCWI data. The images of the source are clearly detected after this subtraction and a spectrum is extracted within a circular aperture of 2 pixels radius centered on each lensed image on the cubes. The four spectra display the same features (although the detection is marginal for the faintest lensed image), confirming the lensed nature of the source. Figure~\ref{fig:spec-source} shows the spectrum which is typical of an ETG and does not display any line emission associated with star formation. A redshift $z=\zsvalue \pm 0.003$ is derived by matching the wavelengths of the main absorption lines (\ion{Ca}{K}, $G$--band, \ion{Mg}{I}\,b, \ion{Na}{I}\,D) to those seen in the data. As evident from Fig.~\ref{fig:spec-source}, those absorption lines are well separated from the spectrum of the lens, further supporting that they are not observational artefacts. 

Quickly after the lens discovery in September 2023, we requested DDT time on the MISTRAL spectrograph mounted on the 1.93m telescope at Observatoire de Haute Provence, which was well suited for this high-latitude target. On December 2nd 2023, we observed the source for 30 minutes with a long slit oriented to maximise the flux coming from the lensed ring. We therefore covered the south east and north west lensed images. No prominent emission line could be found, but an indication of a low S/N emission line near $7860\,\AA$ was observed, suggesting a source redshift of $\zs\sim1.1$. This was subsequently disproved by the deeper KCWI IFU data described previously. Reconsidering the MISTRAL spectrum with the knowledge of the lensed source redshift, no significant absorption lines were visible in the north west lensed image. We did however find an isolated absorption line with $\mathrm{S/N}\ge$2 in the south east image which could correspond to the K line at $\zs=0.4057\pm0.0001$.

\subsection{Strong lens modelling}
\label{sec:lens-modelling}
Using the VIS \IE image with the light model described in Sect.~\ref{sec:lens-light-distribution} and Fig.~\ref{fig:vis-light-model} subtracted, we model the strongly lensed emission using the {\tt pronto} software \citep{Vegetti2009,Rybak2015b,Rizzo2018,Ritondale2019,Powell2021,Ndiritu2024}. We briefly summarise the method here but the cited papers should be consulted for more details.

The lens galaxy total mass-density distribution is described by an elliptical power-law \citep{Barkana1998,Tessore2015}, with Einstein radius, $\erad$, axis ratio, $q$, position angle, $\posa$ and slope, $\gamma$. We also test models that add an external shear component, and the boxiness already observed in the light profile (see Fig.~\ref{fig:isophote-properties}), modelled as an $m=4$ multipole perturbation formulated as in \citet{ORiordan2024}. We use a Gaussian prior of mean zero and standard deviation $1\%$ for the multipole amplitudes. The posterior distribution of the non-linear model parameters is found using the \texttt{MultiNest} nested sampler which also returns the Bayesian evidence, $\ev$, for a given combination of model and data \citep{Feroz2009}. The source surface brightness distribution is reconstructed on an irregular Delaunay grid in the source plane. For each set of non-linear model parameters, the best-fit source is found by solving a linear system which includes the image plane brightnesses, the deflection angle in each pixel, and the PSF blurring. A regularisation scheme is applied that minimises local curvature in the source plane surface brightness. The strength of this regularisation, $\lambdas$, is a free parameter. Larger values of $\lambdas$ increase the Bayesian evidence, so that simpler sources are favoured over complex ones, implicitly accounting for Occam's Razor.

During modelling, we make two modifications to the typical method used in the other {\tt pronto} papers cited above. These are necessary to deal with the very high S/N in the lensed images. First, the linear system used to find the source surface brightness only includes eight out of every nine pixels. In other words, in each $3\times3$ square of pixels, all but the pixel in the top-left corner are solved for as described above, with the brightness in the corner pixel interpolated from its neighbours. This is done to prevent overfitting, which occurs when all pixels are included, because the dynamic range in the source plane is large enough to allow all image plane noise to be absorbed in the source model for low values of $\lambdas$. Second, the regularisation strength, $\lambdas$, is multiplied by a per-pixel weighting factor that depends exponentially on the S/N in each pixel. This mitigates overfitting but, more importantly, enforces smoothness on the low-surface brightness ring and allows for sharper gradients at the centre where the source is cuspy.

In all models, we fit the data within a mask enclosing the Einstein ring and excluding pixels belonging to any of the bright compact objects revealed in the lens light subtraction. One compact object, highlighted in Fig.~\ref{fig:vis-light-model}, is too close to one of the lensed images to be masked, so we fit a circular S\'ersic profile to its emission in the image plane simultaneously with the strong lens modelling described above. We also test a model that includes a mass component for this object, but it is not preferred by the data ($\Delta \ln \ev=-4.5$ against the first model in Table~\ref{tab:modelling-results}), so only its light is modelled. Assuming that the object is at the redshift of the lens, its absolute magnitude is $M\left(\IE\right)=-11.8$, and the inferred S\'ersic profile effective radius is $\Reff=48^{+4}_{-3}\,\mathrm{pc}$ which is consistent with a small globular cluster \cite[see e.g.][]{He2018}.

\begin{table*}[]
	\caption{Comparison of strong lensing models}
	\centering
	\begin{tabular}{ r r r r r r r r r }
		
		Additions & $\erad$ & $\gamma$ & $q$ & $\posa$ & $\abs{\gamma_\mathrm{ext}}$ & $m_4$ &$\lambdas$ & $\Delta \ln \ev$\\
		\hline
		$\gamma_\mathrm{ext}$+$m_4$ & $\ang{;;2.500}$ & 2.03 & 0.702 & 143.3 & 0.027 & 0.012 & 203 & 133.3\\
		$\gamma_\mathrm{ext}$+$m_4$+GC & $\ang{;;2.500}$ & 2.13 & 0.684 & 143.3 & 0.019 & 0.018 & 198 & 128.9\\
		$\gamma_\mathrm{ext}$ & $\ang{;;2.503}$ & 1.97 & 0.714 & 143.7 & 0.032 & - & 181 & 95.9\\
		$m_4$ & $\ang{;;2.500}$ & 2.33 & 0.637 & 144.6 & - & 0.037 & 175 & 78.6\\
		None & $\ang{;;2.496}$ & 2.26 & 0.698 & 144.6 & - & - & 148 & 0.0\\
		\hline
	\end{tabular}
	\tablefoot{The deflector is an elliptical power-law (EPL) to which we add either an external shear, $\extshear$, or an $m=4$ multipole representing boxiness, or both. One model includes both angular additions plus a mass component for the assumed globular cluster (GC). The columns give the maximum a posteriori values for: the Einstein radius, $\erad$, the power-law slope, $\gamma$, the axis ratio, $q$, the position angle, $\posa$, the absolute strength of the external shear component, $\abs{\gamma_\mathrm{ext}}$, the strength of the $m=4$ multipole, $m_4$, the source regularisation strength $\lambdas$, and the difference in log-evidence versus the simplest model. The number of significant figures given indicates the mean uncertainty on that parameter across the different models (see also Sect.~\ref{sec:strong-lens-discussion}).}
	\label{tab:modelling-results}
\end{table*}

\begin{table}[]
	\caption{Lensed image properties}
	\centering
	\begin{tabular}{ p{2cm} r r r }
		
		Image & $\theta_x$ & $\theta_y$ & $\mu$\\
		\hline
		1 & $\ang{;;1.55}$ & $\ang{;;1.93}$ & 14.3\\
		2 & $\ang{;;2.37}$ & $\ang{;;-0.79}$ & 10.1\\
		3 & $\ang{;;-1.43}$ & $\ang{;;-1.88}$ & 13.0\\
		4 & $\ang{;;-1.57}$ & $\ang{;;2.06}$ & 9.9\\
		\hline
	\end{tabular}
	\tablefoot{The angular $x$ and $y$ positions and the scalar magnification $\mu$ for each lensed image. The images are numbered in the clockwise direction starting with the brightest image, as in Fig.~\ref{fig:lens-modelling}, frame A. The angular coordinates are given relative to the centre of the lens at \galaxycoords. Magnification values are given to a fiducial precision of one decimal place, discussed in Sect.~\ref{sec:strong-lens-discussion}.}
	\label{tab:image-properties}
\end{table}

The modelling results are printed in Table~\ref{tab:modelling-results}, and the best model is shown in Fig.~\ref{fig:lens-modelling}. The $\gamma_\mathrm{ext}$+$m_4$ model, including both external shear and boxiness in the mass distribution, returns the largest value of $\ev$. It also has the smoothest source reconstruction, indicated by the largest $\lambdas$ value. The amplitude of the boxiness found in the mass model is $1.24^{+0.14}_{-0.08}\%$. This is similar to the amplitude measured in the isophotes in the region where we expect strong lensing to be sensitive to the mass distribution, that is, at and just inside the Einstein radius (see Fig.~\ref{fig:isophote-properties}). This would suggest that the total mass and light distributions share similar morphologies in the inner region of \galaxyname.

We measure the magnification of the lensed images by casting a quadrilateral enclosing each image's brightest pixel to the source plane, using the best lens model found here. The ratio of the pixel area in the source and lens planes gives a point estimate of the magnification. This is printed in Table~\ref{tab:image-properties} with the relative angular positions of the four images, measured by fitting the PSF kernel to each image in the image plane. Although this is a reliable way to find the centroid of each image, the PSF proves to be a poor fit to the profile of the lensed images, indicating that there is no strong point-like emission in the images. The total magnitude of the strongly lensed emission inside the mask shown in Fig.~\ref{fig:lens-modelling}, panel A, is $\IE=18.1$. The total magnitude of the unlensed emission in the source plane, using the $\gamma_\mathrm{ext}$+$m_4$ lens model presented here is $\IE=21.3$.

\subsection{Stellar initial mass function and dark matter content}
\label{sec:imf}

We combine the stellar velocity dispersion from the DESI spectrum with a measurement of the Einstein radius to construct a simple two-component model of the central region of \galaxyname. We assume a spherical NFW halo whose normalisation we infer, using a Gaussian prior centred on $M_{500} = 4.1 \times 10^{13}\Msun$, as measured with the SZ effect by \cite{Pratt2020}, and a prior width given by the measurement uncertainty on $M_{500}$ \citep[0.14 dex,][]{Kettula2015}. We assume a concentration $c_{500}=3.27$ ($c_{200}=4.96$ for $M_{200} = 5.5 \times 10^{13}\Msun$) given by the halo-concentration relation from \cite{Ishiyama2021}. We use a multi-Gaussian decomposition of the lens light profile which yields a measurement of the effective radius, $\Reff \sim \ang{;;12.3}$. We allow the stellar mass-to-light ratio (constant across the galaxy) to be a free parameter, along with a constant anisotropy parameter $\beta = 1 - (\sigma_\mathrm{t}/\sigma_\mathrm{r})^2$ where $\sigma_\mathrm{t}$ and $\sigma_\mathrm{r}$ are the tangential and radial anisotropies respectively. We calculate model velocity dispersions within DESI's $\ang{;;1.5}$ aperture using the Jeans Anisotropic Modelling (JAM) code from \cite{JAM}. We construct a simple $\chi^2$ likelihood comparing the model predictions of the velocity dispersion and Einstein radius with our measurements. We then sample the posterior probability distribution using \texttt{emcee} \citep{emcee}. We convert the inference on the stellar mass-to-light ratio to an IMF mismatch parameter as
\newcommand{\mstarmod}{M_{\star}^\mathrm{mod}}
\begin{equation}
	\imfmismatch = \frac{\mstarmod}{\mstarChab}\,,
\end{equation}

\begin{figure}
	\centering
	\includegraphics[width=1.0\columnwidth]{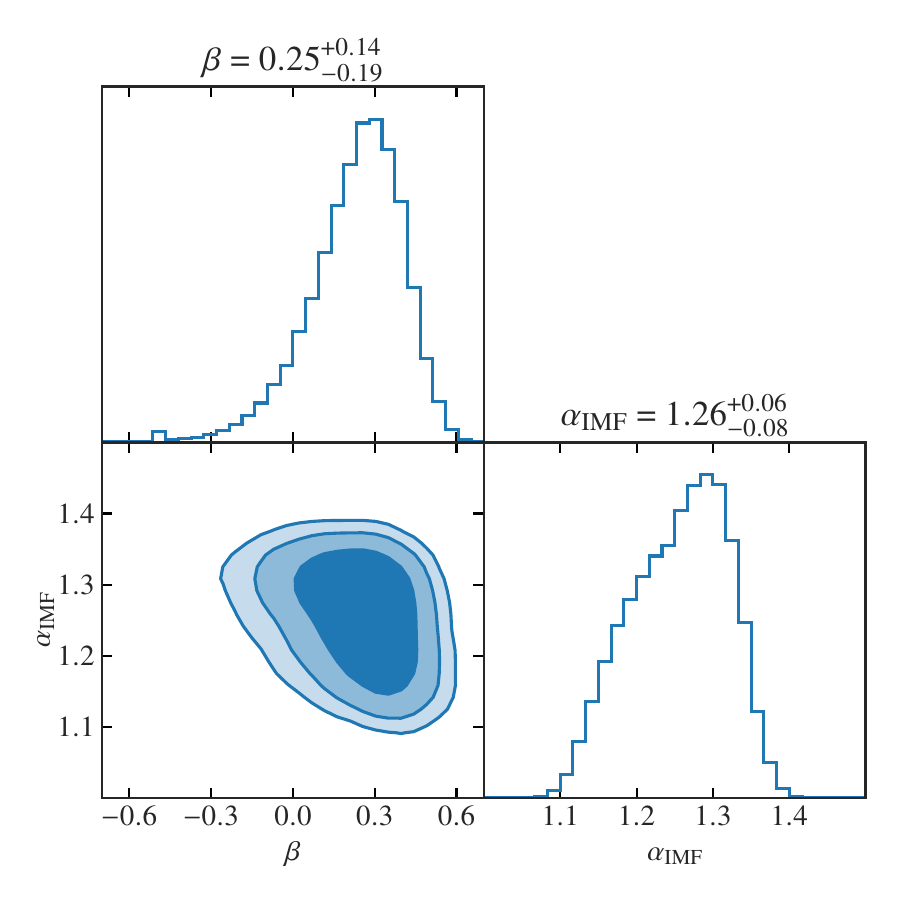}
	\caption{Posterior distributions for the velocity anisotropy parameter, $\beta$, and IMF mismatch parameter, $\imfmismatch$, from combining the DESI velocity dispersion with the Einstein radius measurement. The model favours a slight radial anisotropy, and an IMF that is heavier than Chabrier ($\imfmismatch = 1$) and lighter than Salpeter ($\imfmismatch \sim 1.7 $).}
	\label{fig:corner-imf}
\end{figure}

\begin{figure}
	\centering
	\includegraphics[width=1.0\columnwidth]{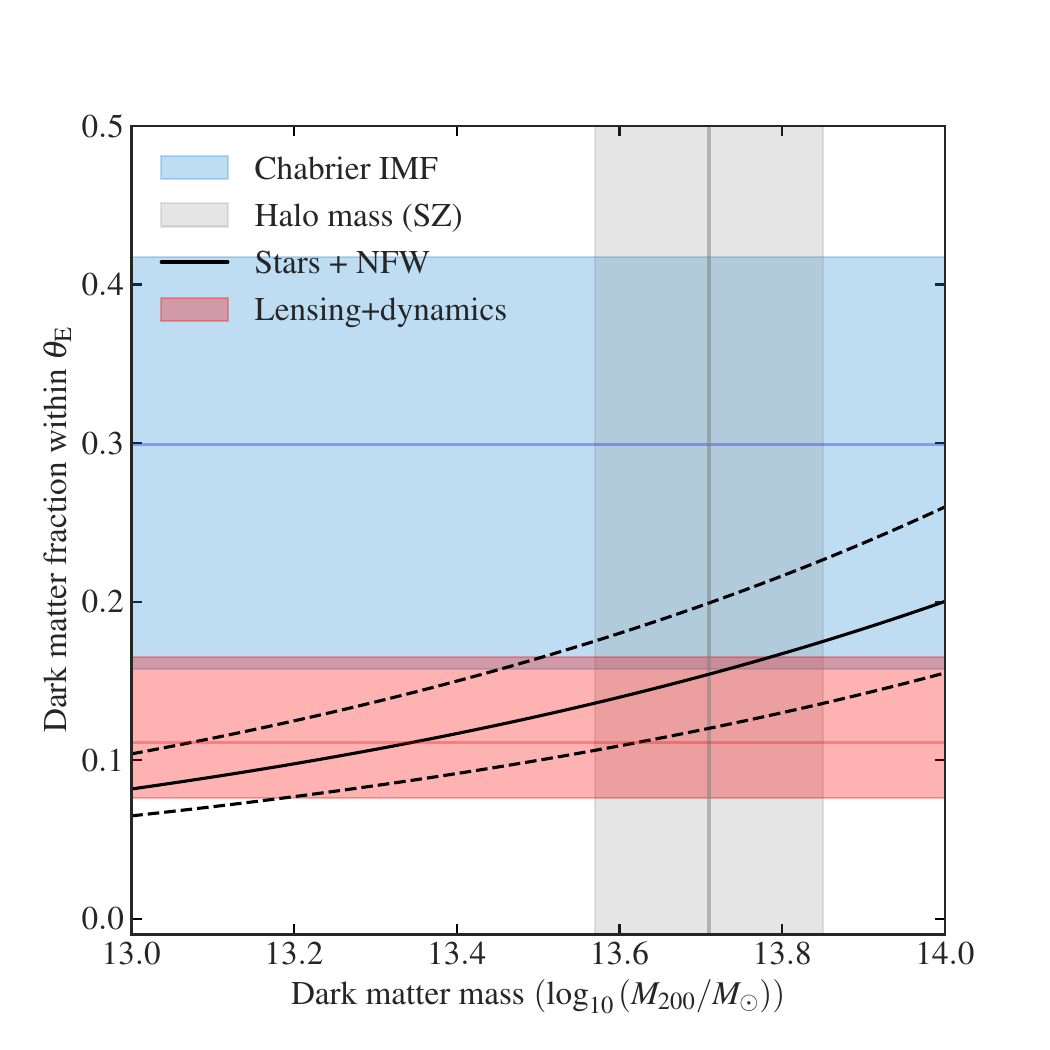}
	\caption{Dark matter fraction within the Einstein radius as a function of halo mass. The solid curve gives the estimated fraction with dotted lines representing the scatter uncertainty in the concentration. The gray region indicates the halo mass from SZ measurement with 1$\sigma$ uncertainty. The blue-shaded area represents the dark matter fraction inferred using the Chabrier IMF and photometric data, while the red-shaded area is derived from lensing and dynamics.}
	\label{fig:darkmatter}
\end{figure}
\noindent
where $\mstarmod$ is the stellar mass within the Einstein radius inferred from the lensing+dynamics model, and $\mstarChab$ is the stellar mass within the Einstein radius measured from SED fitting in Sect.~\ref{sec:lens-kinematics}.

We measure a projected stellar mass within the Einstein radius of $\mstarmod = (1.36_{-0.09}^{+0.06}) \times 10^{11}\Msun$, and an anisotropy $\beta = 0.26_{-0.18}^{+0.14}$. This gives a projected dark matter fraction within the Einstein radius, $\dmfraction = (11.1_{-3.5}^{+5.4})\%$. The constraint on the stellar mass translates to an inference on the IMF mismatch parameter, $\imfmismatch = 1.26_{-0.08}^{+0.05}$. Figure~\ref{fig:corner-imf} shows our inference on the anisotropy and IMF mismatch parameter, $\imfmismatch$, of the lens. We note that this implies an IMF in the central regions of \galaxyname~that is heavier than Chabrier but lighter than Salpeter.

Figure \ref{fig:darkmatter} compares the projected dark matter fraction within the Einstein radius as calculated from the SED modelling, which assumes a Chabrier IMF, and mass modelling inferred from combining the strong lensing and dynamics, which makes no assumptions about the IMF. This is compared to the dark matter fraction as a function of NFW halo mass, given that the mass enclosed by the Einstein radius is $1.53 \times 10^{11} \Msun$ and accounting for an uncertainty of $\pm$ 0.14 dex in halo concentration. We note that the $(M-\sigma)$ relation from \cite{gultekin2009ApJ...698..198G} implies a central SMBH mass of $\logten\left(M_\mathrm{BH}/\Msun\right)$ = 8.95 $\pm$ 0.34. Since this corresponds to only $(0.57^{+0.68}_{-0.31})\%$ of the mass within the Einstein radius, we do not include it in our analysis.

\section{\label{sec:discussion}Discussion}

\subsection{Serendipity of discovery}

Low redshift lenses are intrinsically rare because there is very little volume at low redshift. Integrating the volumetric velocity dispersion function measured in SDSS \citep{Choi2007} implies that there should only be $\sim 2400$ galaxies with $\sigmav > 250\,\kms$ and $z < 0.05$. That we observed one in the early days of \Euclid is unremarkable, but for it to be an obvious strong lens is quite exceptional. Using the lens population model of \citet{Collett2015}, we find that an isothermal $300\,\kms$ galaxy at $z=0.04$ has only a 1 in $2000$ chance of producing a strong lens system with $\IE<19$. \galaxyname~has a 1 in 30 (1 in 6) chance of multiply imaging a source to brighter than 22nd (24th) magnitude in $\IE$. Such faint arc systems would be much harder to detect serendipitously.

The exceptional nature of \lensname~means it is unlikely that \Euclid will find another lens below $z=0.05$ with a ring as bright as that observed here~$(P=0.01)$. However, \Euclid's $14\,000\,\deg^2$ dataset should contain 4 (20) lenses below $z=0.05$ with ring magnitudes brighter than 22nd (24th) magnitude in $\IE$.\footnote{These forecasts are for lenses with a velocity dispersion, $\sigmav> 250\,\kms$, a minimum Einstein radius of $\ang{;;0.5}$ and a total source magnification of at least 3.} These numbers are increased by a factor of 9 if lenses below $z=0.1$ are considered low redshift. \Euclid should therefore increase the number of known low redshift lenses by at least a factor of $\sim 5$.

\subsection{Peculiar velocity of \galaxyname}

Since our Universe is expanding, most galaxies are moving away from us and we observe their spectra to be redshifted. In the nearby Universe, the cosmological redshift is proportional to the distance to the galaxy. However, galaxies are also moving relative to the Hubble flow. The observed redshift is a mix of the cosmological redshift and the Doppler shift resulting from the galaxy's peculiar velocity. Much of the peculiar velocity of galaxies comes from large scale bulk flows as matter flows towards superclusters and away from voids. The recession velocity of \galaxyname~is approximately $1.3 \times 10^4$ km s$^{-1}$. Most of this is due to the expansion of the Universe, but the recession velocity is not so great that the peculiar velocity of \galaxyname can be completely ignored. The low-redshift cosmic flow model of \citet{Kourkchi2020} suggests that $3\%$ of the redshift of \galaxyname~comes from a bulk flow away from us. On top of the bulk flow peculiar velocities of individual galaxies are typically present at the $150\,\kms$ level. Combining these two effects results in a cosmological redshift of $z=0.041\pm0.001$ for \galaxyname.

\subsection{Strong lens modelling considerations}
\label{sec:strong-lens-discussion}
Our modelling results presented in Table~\ref{tab:modelling-results} show a preference in the data for an external shear component in the lens model, i.e, any model with $\gamma_\mathrm{ext}$. We attribute this to a lack of appropriate model complexity, rather than the shearing effect of a nearby object. This is for three reasons. First, the measured strength of the external shear component is small ($2.7\%$) and its orientation ($49$~$\deg$) is almost exactly that of the semi-minor axis of the lens. These are conditions identified by \citet{Etherington2024} in other spurious external shear measurements. Second, in the models that use external shear, the axis ratio (ellipticity) of the lens is increased (decreased), suggesting a degeneracy between the two. Third, there is no obvious culprit in the vicinity of the lens for such shear; if the source of the shear were on the lens plane it would require substantial density. It should be noted though, that \citet{Henry1995} find evidence in X-ray emission that the galaxy is part of a small group.

The isophote properties in Fig.~\ref{fig:isophote-properties} show significant changes in the angular structure of the stellar component within the Einstein radius. This cannot be fully accounted for by the simple power-law plus multipole (or shear) model used here. Nevertheless, the measurement of the Einstein radius should be robust, as evident in the similarity of the inferred values in Table~\ref{tab:modelling-results}, despite the very different angular structure of the models. The inferred values of the axis ratios between mass in Table~\ref{tab:modelling-results} and light in Fig.~\ref{fig:isophote-properties} also differ significantly, and this depends on whether shear or multipoles or both are included. Similarity between the two should be expected only if the dark-matter contribution inside the Einstein radius is small, and if its morphology is similar to that of the stellar component. In this case the former is likely true, based on the analysis of Sec~\ref{sec:imf}. To determine if the latter is true, a joint mass-light model should be used on the unsubtracted data. If the dark matter component is, for example, more elliptical than the stellar component, it would explain the more elliptical model inferred by the strong lens modelling compared to the isophotes.

It should be noted that in Tables~\ref{tab:modelling-results}~and~\ref{tab:image-properties} values for the model parameters and magnifications are given to a precision which is indicative of the uncertainty on those values. Exact values for the uncertainty are not meaningful in this case. This is because the extreme S/N of the observation means the error budget is dominated by the systematic uncertainty intrinsic to the strong lens modelling. It is well known that models which present a poor fit to strong lensing data can be constrained with high precision. The uncertainties indicated here should therefore be considered lower-bounds which only account for measurement uncertainty, and include no contribution from the systematic uncertainty. In future works, more complex and appropriate models will be used which also incorporate information from the stellar light distribution and dynamics. This can mitigate the systematic uncertainties in the strong lensing-only modelling work. We will also assess the lens modelling systematics in a similar manner to \citet{Galan2024}. In that work, it was also shown that a sub-pixel PSF model, as opposed to the pixel-level PSF used here, can mitigate systematic uncertainties on the lens parameters when the source is cuspy.

\subsection{IMF measurement}

With a simple combination of lensing and dynamical constraints, we infer a stellar mass-to-light ratio suggesting an IMF heavier than Chabrier, and lighter than Salpeter, $\imfmismatch = 1.26_{-0.08}^{+0.05}$. \cite{Newman2017} measured $\imfmismatch$ for two low-redshift (SNELLS) lenses using lensing and dynamics. They similarly found mismatch parameters lighter than Salpeter. In contrast, measurements at higher redshift, such as SLACS \citep{Auger2010, Treu2010}, infer an IMF consistent with Salpeter. Since low-redshift lenses only probe the IMF in the very central regions ($\erad/R_\mathrm{eff} \sim 0.2$ in \lensname), the IMF measurement presented here underlines the challenge that these low-redshift lenses present to the currently favoured scenario in which the IMF varies radially and is more bottom-heavy in the centres of ETGs than in the outskirts. 

It is important to note that our model of the dark matter halo is limited to a spherical NFW profile with a fixed concentration (see Sect.~\ref{sec:imf}) while more flexible models are possible. We make this choice due to the lack of mass constraints at large radii. However, it is possible that not allowing full flexibility may bias the inference of the IMF mismatch parameter. Using a generalised NFW where the inner slope is a free parameter would especially help to mitigate the mass-anisotropy degeneracy. We leave to a future paper the task of jointly inferring a more flexible dark matter halo mass using 2D lensing and kinematic information. Given the $\sim 2\sigma$ tension between the IMF inferred from lensing and dynamics, with that inferred from stellar population modelling reported in \cite{Newman2017}, it will also be important to compare these results with a full stellar population model of the spectrum within the Einstein radius.

\section{Summary}
\label{sec:summary}

\Euclid is forecast to discover $10^5$ new strong gravitational lenses and, alongside other surveys, usher in a new era of big data for the field. In this paper we present imaging and spectroscopic data for the very first of these new strong lenses. A complete Einstein ring ($\zs=\zsvalue$) was discovered around the nearby galaxy \galaxyname ($\zl=0.042$) which was identified in \Euclid imaging for the first time, lensing a compact object at $\zs=\zsvalue$. 

The lens is significant also for the very low redshift of the deflector, with only $\sim 5$ other lenses known at such a distance. Strong lenses at low redshift have Einstein radii that are comparatively small in physical terms and allow for a detailed study of the composition and structure of the central region of the galaxy. In this paper we presented initial results from these studies. As well as imaging data from \Euclid VIS and NISP, we also obtained resolved spectroscopy from KCWI and spectroscopy for the central region of the galaxy from DESI.

From the DESI spectrum we inferred a central velocity dispersion of $\sigmav=303\pm15\,\kms$. We modelled the light profile of the lens galaxy in detail, finding a complex and varying isophotal structure inside the Einstein ring. Subtracting this light model from the VIS data, we modelled the strongly lensed emission using a pixellated source reconstruction and an elliptical power-law mass model. We measured an Einstein radius of $\erad=\ang{;;2.500}\pm\ang{;;0.001}$ and a strong preference ($\Delta \log \ev=37.4$) for boxiness in the total mass profile, consistent with that in the light.

With the Einstein radius inferred from strong lens modelling and the velocity dispersion inferred from the DESI spectrum, we used JAM to infer the properties of the central region of \galaxyname. We found a heavier-than-Chabrier IMF in the central region, with mismatch parameter $\imfmismatch = 1.26_{-0.08}^{+0.05}$, and a dark matter fraction inside the Einstein radius of $\dmfraction = (11.1_{-3.5}^{+5.4})\%$.

Simulations indicate that the discovery of this object was highly improbable, with such a large galaxy at this redshift having only a $1$ in $2000$ chance to lens a source as bright as that observed here. Between 4 and 20 low-redshift lenses should remain to be discovered in the \Euclid survey area, although typically with much dimmer sources. We propose the nickname `Altieri's lens' for the object, in recognition of the incredible fortune of its discoverer, B. Altieri, when inspecting early \Euclid VIS data in September 2023.

\begin{acknowledgements}
\AckERO
\AckEC  

C.~O’R, S.~V., and D.~P. thank the Max Planck Society for support through a Max Planck Lise Meitner Group.

A.~N. gratefully acknowledges the support of the Belgian Federal Science Policy Office (BELSPO) for the provision of financial support in the framework of the PRODEX Programme of the European Space Agency (ESA) under contract number 4000143347.

This project has received funding from the European Research Council (ERC)
under the European Union’s Horizon 2020 research and innovation
programme (LensEra: grant agreement No 945536).

T.~E.~C. is funded by a Royal Society University Research Fellowship. 

Based in part on observations made at Observatoire de Haute Provence (CNRS), France, with MISTRAL. This research has made use of the MISTRAL database, operated at CeSAM (LAM), Marseille, France.

Some of the data presented herein were obtained at Keck Observatory, which is a private 501(c)3 non-profit organization operated as a scientific partnership among the California Institute of Technology, the University of California, and the National Aeronautics and Space Administration. The Observatory was made possible by the generous financial support of the W. M. Keck Foundation. The authors wish to recognize and acknowledge the very significant cultural role and reverence that the summit of Maunakea has always had within the Native Hawaiian community. We are most fortunate to have the opportunity to conduct observations from this mountain.

DESI construction and operations is managed by the Lawrence Berkeley National Laboratory. This research is supported by the U.S. Department of Energy, Office of Science, Office of High-Energy Physics, under Contract No. DE–AC02–05CH11231, and by the National Energy Research Scientific Computing Center, a DOE Office of Science User Facility under the same contract. Additional support for DESI is provided by the U.S. National Science Foundation, Division of Astronomical Sciences under Contract No. AST-0950945 to the NSF’s National Optical-Infrared Astronomy Research Laboratory; the Science and Technology Facilities Council of the United Kingdom; the Gordon and Betty Moore Foundation; the Heising-Simons Foundation; the French Alternative Energies and Atomic Energy Commission (CEA); the National Council of Science and Technology of Mexico (CONACYT); the Ministry of Science and Innovation of Spain, and by the DESI Member Institutions. The DESI collaboration is honored to be permitted to conduct astronomical research on Iolkam Du’ag (Kitt Peak), a mountain with particular significance to the Tohono O’odham Nation.

\end{acknowledgements}

\bibliography{bibliography}

\end{document}